\documentclass[preprint,12pt]{elsarticle}

\usepackage{amsmath,amsfonts}
\usepackage{array}
\usepackage[caption=false,font=footnotesize,labelfont=rm,textfont=rm]{subfig}
\usepackage{textcomp}
\usepackage{stfloats}
\usepackage{url}
\usepackage{verbatim}
\usepackage{graphicx}

\usepackage{amssymb}

\usepackage{graphicx}
\usepackage{graphics}
\usepackage{epsfig}
\usepackage{paralist}
\usepackage{svg}
\usepackage{amsthm}
\usepackage{amsmath}
\usepackage{autobreak}
\usepackage{float}                  
\usepackage{subfig}                 
\usepackage{overpic}
\usepackage{lipsum}
\usepackage{cuted}
\usepackage{amsfonts}
\usepackage{mathrsfs}
\usepackage{enumitem}
\usepackage{booktabs} 
\usepackage{array}
\usepackage{tabularx}
\usepackage[ruled,linesnumbered]{algorithm2e}
\usepackage{amssymb}

\newtheorem{definition}{Definition}

\journal{Computer Networks}

\begin{document}

\begin{frontmatter}



\title{HL-DPoS: An Enhanced Anti-Long-Range Attack DPoS Algorithm}


\author[1]{Yang Li}
\ead{johnli@buaa.edu.cn}
\author[2,3]{Chunhe Xia}
\author[4]{Chunyan Li}
\author[1]{Yuan Zhao}
\author[1]{Chen Chen}
\author[4]{Tianbo Wang\corref{cor1}}
 \ead{wangtb@buaa.edu.cn} 
  \cortext[cor1]{Corresponding Author}
  
\affiliation[1]{organization={School of Computer Science and Engineering, Beihang University},
            city={Beijing},
            country={China}}

\affiliation[2]{organization={Beijing Key Lab. of Network Technology, Beihang University},
            city={Beijing},
            country={China}}
            
\affiliation[3]{organization={Guangxi Collaborative Innovation Center of Multi-source Information Integration and Intelligent Processing, Guangxi Normal University},
            city={Guilin},
            country={China}}
            
\affiliation[4]{organization={School of Cyber Science and Technology, Beihang University},
            city={Beijing},
            country={China}}
\begin{abstract}
The consensus algorithm is crucial in blockchain for ensuring the validity and security of transactions across the decentralized network. However, achieving consensus among nodes and packaging blocks in blockchain networks is a complex task that requires efficient and secure consensus algorithms. The DPoS consensus algorithm has emerged as a popular choice due to its fast transaction processing and high throughput. Despite these advantages, the algorithm still suffers from weaknesses such as centralization and vulnerability to long-range attacks, which can compromise the integrity of the blockchain network.
 To combat these problems, we developed an En\underline{h}anced Anti-\underline{L}ong-Range Attack DPoS algorithm (HL-DPoS). First, we split nodes into pieces to reduce centralization issues while giving witness nodes the power to report and benefit from malicious node's reports, maintaining high efficiency and high security. Second, we propose a validation method in HL-DPoS that compares consensuses transactions with the longest chain to detect long-range attacks. Algorithm analysis and simulation experiment results demonstrate that our HL-DPoS consensus algorithm improves security while achieving better consensus performance.  
\end{abstract}




\end{frontmatter}



\section{Introduction}

With people's increasing concern about information security and privacy protection, blockchain technology has received widespread attention this year, which integrates multiple network security technologies to help the Internet realize the evolution from digit information to real value. Blockchain can be viewed as a decentralized public ledger, and transmission of data between nodes in the blockchain network is accomplished through transactions\cite{Nofer2020Blockchain, Lu2020BlockchainAF}. 
The consensus mechanism is the core of blockchain technology, ensures the consistency and determinacy of the block content stored by each node, enabling all nodes in the blockchain network to collaborate\cite{Xu2023ASO, Wang2023AFA}
organizes nodes to package blocks in chronological order, ensuring that all nodes reach consensus on the blockchain in an untrusted and distributed environment. Nodes participating in block packaging will be rewarded upon completion to improve the enthusiasm of nodes participating in the packaging process.

The DPoS consensus algorithm selects a queue of witness nodes by allowing nodes in the blockchain network to vote for each other; it's a voting-based consensus representing democracy and efficiency. All nodes in the blockchain vote for their trust node and select a certain number of nodes, for example, 101, as all nodes represent pack blocks in turn, called witness nodes or package nodes. Every node can elect its trust node to be the witness node, or itself can be elected, so every node's opinion can be expressed, and the witness nodes who finally enter the package queue represents the choice of most nodes. Only the witness nodes have the power to package blocks. DPoS reduces block package power competition progress, which node should package block has been specified at every round beginning, which makes block package highly efficient \cite{Kaur2021ARS,Wan2019RecentAI,Liu2022ResearchOD}.

The advantage of the DPoS consensus algorithm is its high efficiency in consensus, reducing computational power consumption compared to PoW consensus algorithm. However, in practical applications, it also has the following issues\cite{Krishnamurthi2021ABA, Rebello2021ASA}:

\begin{enumerate}

\item Centralization Issue.

The block is consensuses by the DPoS consensus algorithm packed by a group of witness nodes elected by the entire network nodes. Some users/entities control multiple blockchain nodes and vote for fixed nodes when voting, resulting in some nodes receiving consistently high votes and a greater chance of entering the witness nodes queue. A few nodes will continuously control the power to pack the blockchain network, leading to centralization issues.

\item Long-Range Attack Issue.

In the witness nodes queue, witness nodes have the power to package blocks, but some nodes may hope to gain greater benefits by launching long-range attacks; the node is considered a malicious node. The long-range attack is an attacking way of creating a longer chain than the main chain to manipulate the transactions in the blockchain.

%
\end{enumerate}

Because the DPoS consensus algorithm generates witness nodes through elections to package blocks, nodes will inevitably join in the package queue repeatedly. The repetition of specific nodes in the package queue will result in predictable witness nodes, the long-term concentration of packaging power in specific nodes, a decrease in the security of the consensus algorithm, and an increased possibility of specific nodes being attacked or launching attacks. In the DPoS consensus algorithm, if some nodes are colluding to vote for a malicious node, the malicious node can launch a long-range attack when packaging blocks. Long-range attacks refer to situations where a malicious node generates a new chain fork from the main chain. Specifically, in the DPoS consensus algorithm, due to the mechanism of accepting the longest chain as the main chain, the fork chain may replace the main chain, and the malicious node will manipulate the transactions saved in the blockchain.


Regarding the problems with the DPoS consensus algorithm, we propose the HL-DPoS consensus algorithm. This algorithm uses verifiable random functions (VRF) \cite{Dodis2005AVR} to shard the nodes in the blockchain, selecting as many different nodes as possible to enter the witness queue, reducing the occurrence of centralization issues. This algorithm gives nodes in the witness nodes queue block to verify and report power, reducing the possibility of witness nodes behaving maliciously. The HL-DPoS consensus algorithm also proposes the longest chain verification mechanism to verify witness nodes' malicious action and solve a technique for long-range attacks, improving the algorithm's security. The main contributions of this paper are summarized as follows:

\begin{itemize}

\item Proposing a novel consensus mechanism as HL-DPoS; the election method of witness nodes is improved in our proposed mechanism to prevent malicious behavior, randomly split all the nodes into groups, choose two nodes from each group as the witness nodes by vote, and random select, makes nodes in witness nodes queue more randomly, the security of HL-DPoS is analyzed in this paper.

\item Proposing the longest chain verification mechanism in HL-DPoS consensus algorithm, where nodes compare the transactions contained in consensuses blocks with blocks in the longest chain when it broadcasts, to check if there have consensuses transactions that don't include in the longest chain, thus detect witness nodes malicious action to preventing long-range attacks.

\end{itemize}

In this article, Section \uppercase\expandafter{\romannumeral2} introduces related works about blockchain consensus algorithms and Nash equilibrium. Section \uppercase\expandafter{\romannumeral3} introduces the Nash equilibrium, verifiable random function (VRF), and proposes a novel consensus algorithm, HL-DPoS, to improve DPoS consensus mechanism security. Section \uppercase\expandafter{\romannumeral4} analyzes the HL-DPoS with Nash equilibrium to explain the security of our proposed mechanism. Section \uppercase\expandafter{\romannumeral5} consists of the experimental simulations of the HL-DPoS. Examples verify the feasibility and effectiveness of a consensus mechanism. Section \uppercase\expandafter{\romannumeral6} concludes this article and assumes the future work.


\section{RELATED WORK}

There have many improvements and applications of different consensus algorithms. Based on the characteristics of decentralization, immutability, and traceability of blockchain technology, Zibin Zheng et al. provide an overview of blockchain architecture, compare some typical consensus algorithms used in different blockchains, list technical challenges, recent progress, and future trends \cite{Zheng2017AnOO}. Yaodong Huang et al. proposes a hybrid consensus blockchain system combining proof of work (PoW) and proof of stake (PoS) to enhance transaction security \cite{Huang2022IncentiveAI}. Tongtong Geng et al. use deep reinforcement learning (DRL) to adjust the blockchain consensus algorithm to intelligent manufacturing applications, establishing a deep reinforcement consensus algorithm that can handle data more densely, with faster calculation speed, higher accuracy, and stronger security, effectively improving the decentralization performance of blockchain systems \cite{Geng2022ApplyingTB}. Zhiming Liu et al. propose the RAFT+ consensus algorithm generated by deep Q-Network (DQN) for block verification nodes based on the distributed consensus algorithm RAFT, which alleviates the instability between different types of IoT terminal devices, allowing these devices to participate in block consensus, maintaining the strong consistency of the blockchain network \cite{Liu2022ADC}. Yanjun Jiang et al. designed and implemented a consensus algorithm called HPBC based on message-passing technology to address the drawbacks of using a Proof of Work (PoW) algorithm in blockchain, such as waste of computing resources, long block confirmation delays, and low throughput, which can work well in an asynchronous network with Byzantine nodes \cite{Jiang2018AHP}. Fan Yang et al. introduced the competition of computing power in PoW into DPoS and designed an improved consensus algorithm called Delegated Proof of Stake With Downgrade (DDPoS) \cite{Yang2019DelegatedPO}. By further modification, the influence of computing resources and interests on block generation is reduced to achieve higher efficiency, fairness, and power delegation in the consensus process, and a downgrading mechanism is proposed to replace malicious nodes and improve security quickly.

There also have consensus algorithms that combine consensus with Nash equilibrium\cite{  Song2022BlockchainsRI, Khezr2022TowardsAT, Zhang2022ABE, Guo2022IncentiveMF, Guo2020BlockchainME}. Charlie Hou et al. proposed SquirRL as a framework that uses deep reinforcement learning to analyze attacks on blockchain incentive mechanisms, and used known attack methods to demonstrate the effectiveness of the framework, including the optimal selfish mining attack in Bitcoin and the Nash equilibrium in block withholding attacks\cite{Hou2021SquirRLAA}. Aggelos Kiayias et al. introduced the first proof-of-stake blockchain protocol, "Ouroboros", which has strict security guarantees and proposes a reward mechanism for proof-of-stake protocols\cite{Kiayias2017OuroborosAP}. Honest behavior is proven to be an approximate Nash equilibrium, neutralizing selfish mining attacks. Jiaxing Qi et al. established a discrete-time non-exhaustive vacation queue with batch service and gated service to study the performance of blockchain systems with light traffic loads. Transactions are treated as arrivals, vacations, and services, and the average response time of transactions is shown to be affected by the transaction arrival rate. Nash equilibrium behavior and social optimization of transactions are studied, and a transaction pricing policy is proposed to maximize social profit\cite{Qi2020NashEA}. Wenbai Li et al. constructed a game model between mining pools based on the PoW consensus algorithm, and analyzed its Nash equilibrium from two perspectives\cite{Li2020MiningPG}. The power of mining pools, the ratio of penetration to be infiltrated, the betrayal rate of dispatching miners in the mining pool, and income selection are discussed, and results are obtained through numerical simulations.

Furthermore, DPoS consensus algorithm is still a mainstream consensus algorithm; projects such as EOS and Bitshares are running now and developing rapidly. But there still needs to be a practical and reliable solution to the centralization and malicious node problems that keeps efficiency with high security in the DPoS consensus algorithm. We proposed our HL-DPoS to solve these two problems, introduced in Section \ref{sec:An improved DPoS Consensus Algorithm}.

\section{HL-DPoS Consensus Algorithm}
\label{sec:An improved DPoS Consensus Algorithm}

In this section, we proposed an improved blockchain consensus algorithm based on the DPoS consensus algorithm, game theory, and verify random function, to solve the centralization problem and prevent long-range attacks.

\subsection{Preliminaries}
\subsubsection{Nash Equilibrium}
Nash equilibrium, also known as non-cooperative game equilibrium, refers to a strategy in which one party in a game process will choose a certain strategy regardless of the opponent's strategy choice, which is called a dominant strategy \cite{Salant1983LossesFH, Maskin1999NashEA, Daskalakis2006TheCO}. If any participant's strategy is optimal under the determination of other participants' strategies, then this strategy combination is called Nash equilibrium. Nash equilibrium can be divided into pure strategy Nash equilibrium and mixed strategy Nash equilibrium.

For a given strategic game, $\Gamma=\left\langle N,(S_i),(u_i)\right\rangle$ and its strategy team $s^\prime=(s_1^\ast,s_2^\ast,\cdots,s_n^\ast)$,

 \begin{equation} u_i (s_i^* ,s_{-i}^* ) \geq u_i ( s_i, s_{-i}^* ),s_i \in S_i , i=1 , 2 , \ldots , n \end{equation}

Hence, $s_i^*$ be a pure strategy Nash equilibrium of $\Gamma$, where $\Gamma$ represents strategic games; $S_i$ represents the strategy set of the $i$th participant, $u_i:S_1 \times S_2 \times \cdots \times S_n \rightarrow R $ represents the set of participants, which is a finite set; $s_i^*$ represents a strategy in the strategy set of participant $i$ that is strictly better than any other strategy, while $s_{-i}^*$ represents the typical strategy set of all other participants except participant $i$.

For a given player $i$ and their pure strategy set $S_i$, a mixed strategy $\sigma_i$ is a probability distribution over $S_i$,  for each pure strategy $s_i \in S_i$, a probability $\sigma_i (S_i)$ is specified as:

 \begin{equation}\sum_{s_i \in S_i}{\sigma_i(s_i)}=1 \end{equation}



\subsubsection{Verifiable Random Function}

In cryptography, verifiable random function (VRF) is a public key pseudorandom function that allows a user with a private key to obtain a verifiable proof and a value \cite{ Dodis2003EfficientCO}. Anyone with the proof and corresponding public key can verify whether the value is correct, but cannot obtain the private key through the proof. In some consensus algorithms used in blockchain, random function used to randomly poll a node for bookkeeping/verification according to certain hashing rules. If any node can reproduce the polling rules, there is a risk of attack on a future bookkeeping/verification node. VRF functions can hide the hashing rules and only disclose the public key and proof for verification during the packaging phase. After packaging is complete, the private key used to generate the proof is disclosed to other nodes to verify the authenticity of the witness node's packaging rights.

The process of VRF are as follows, input $(\iota)$ and output$(\iota)$ are polynomials in security parameter $\iota$:

\begin{itemize}

\item Generator function $(G(\iota))$: The generator function receives a string $\iota$ as a security parameter, outputs two binary strings, public key $\mathbb{PK}$ and security key $\mathbb{SK}$.

\item Evaluator function$(E(SK,x))$: The evaluator function receives two strings as input, security key $\mathbb{SK}$ and string $x$, use two functions to calculate the result and proof of $\mathbb{SK}$ and $x$: $result=Function_1(SK,x)$ \&  $proof = Function_2 (SK,x)$, $result \in \{0,1\}^{output(\iota)}$.

\item Verify function$(V(\mathbb{PK},x,result,proof))$: The verify function receives public key $\mathbb{PK}$, $result$, $x$, and $proof$ as input parameters, calculate by verify function, output a result $bool$ to show if the given result is correct, $ bool \in\{0,1\}$.
\end{itemize}

VRF has three features: Provability, Uniqueness, and Pseudo randomness. Given functions $input(\iota)$ and $output(\iota)$ are polynomial in given $\iota$.
\begin{itemize}

\item Provability. For any pairs $(\mathbb{PK},SK) \in G(\iota)$ and any input string $x \in \{0,1 \}^{input (\iota)}$, if $(result,proof)=E(SK,x)$, there has a negligible polynomial function $\omega(\iota)$ that:
 \begin{equation}probability\left[V(\mathbb{PK},x,result,proof)=YES\right]>1-2^{-\omega(\iota)}\end{equation}

\item Uniqueness. For any pairs $(\mathbb{PK},SK) \in G(\iota)$ and any input string $x \in \{0,1 \}^{input(\iota)}$, there is no tuple ( $result_1$ , $result_2$ , $\cdots$ , $result_{\mu}$ , $\cdots$ , $result_{\chi}$ , $proof_1$ , $proof_2$ , $\cdots$ , $proof_{\mu}$ , $\cdots$ , $proof_{\chi}$ ) makes Equation \ref{probability-1}.

 \begin{equation}
\begin{aligned}	
& probability \left[result_{\mu} \neq result_{\chi} \middle|\begin{pmatrix}{V(\mathbb{PK},x,result_{\mu},proof_{\mu})=YES}\\
{V(\mathbb{PK},x,result_{\chi},proof_{\chi})=YES}
\end{pmatrix}\right] \\
& <2^{-\omega(\iota)}
\end{aligned}
\label{probability-1}
 \end{equation}


\item Pseudo randomness. For any probabilistic polynomial time distinguishers PD, there has a negligible polynomial function $ \omega (\iota)$ makes Equation \ref{probability-2}.

%
%

\begin{equation}
\left|
\begin{aligned}
probability \left[PD(G(\iota),Function_1(SK,x))=1\right]
\\ -probability \left[PD(G(\iota),\{0,1\}^{output(\iota)})=1\right]
\end{aligned}
\right| \le \omega (\iota) 
\label{probability-2}
\end{equation}

\end{itemize}

\subsection{HL-DPoS Consensus Algorithm}

\subsubsection{Election strategy in HL-DPoS}

\begin{definition}
The HL-DPoS Consensus Algorithm $ \delta$ = $\langle$ $D_l$, $\mathbb{PL}_{\varepsilon}$, $(A_i)_{i \in \mathbb{N}}$, $T$, $P$, $(I_j)_{j\in \mathbb{N}}$, $(u_k)_{k\in \mathbb{N}}$ $\rangle$, $\mathbb{N}$ is a finite set.
\label{HL-DPoS Consensus Algorithm}
\end{definition}

\begin{itemize}

\item $D_l(l=1,2,...,n_1)$ is the set of participating nodes, $n_1 \in \mathbb{Z}^+$ is participating nodes number;
\item $\mathbb{PL}_{\varepsilon}, \varepsilon < n_1 \cap \varepsilon \in \mathbb{Z}^+$, is the set of witness nodes queue;
\item $A_i (i = 1,2, \ldots, n_2)$ is the action set of $\mathbb{PL}_{\varepsilon}$, $n_2 \in \mathbb{Z}^+$ is the action sets' number;
\item T is a set composed of all terminal histories, where a terminal history is an action path from the root node to a terminal node that is not a proper sub-history of any other terminal history. The set consisting of all proper sub-histories (including the empty history $\eta$) of all terminal histories is denoted by cap $S$ and sub cap $T$;
\item $P: S_T \rightarrow \mathbb{PL}_{\varepsilon}$ is a player function, it associates each sub-history with the corresponding participating node;
\item $I_j(j=1, 2, \ldots, n_3)$  is a collection composed of all information sets of witness node $\mathbb{PL}_{\varepsilon}$;
\item Each termination history of the witness node $\mathbb{PL}_{\varepsilon}$ is provided with utility $u_k:T \rightarrow \mathbb{R} (k=1, 2,\ldots, n_2)$, $\mathbb{R}$ is real number set.
\end{itemize}

\begin{table}[htbp]
  \centering
  \caption{Notations in This Paper}
  \label{tab:Notations in This Paper}
  \begin{tabularx}{\columnwidth}{lX}
    \toprule
    Notations & Descriptions  \\
    \midrule
    $\kappa$ & security parameter  \\
    $\mathbb{PK}$ & VRF public key  \\
    $\mathbb{SK}$ & VRF private key \\
    $\mathbb{X}$ & VRF input string \\
    $s$ & VRF calculate result \\
    $\mathbb{V}$ & random value \\
    $\mathbb{P}$ & random point on elliptic curve \\
    $\mathbb{H}$ & hash function \\
    $\mathbb{G}$ & nodes group \\
    $D_l$ & blockchain node \\
    $\mathbb{G}_D^i$ & node $D_l$ in $i$th group \\
    $\mathbb{RL}^i$ & $i$th group nodes votes ordered list \\
     $\psi$ & round limits to enter package list \\
    $\mathbb{NUM}$ & nodes number \\
    $\mathbb{CD}$ & candidate queue \\
    $\mathbb{R}$ & real number set \\
    $R_v^i$ & $i$th group represent node by voting\\
    $R_s^i$ & $i$th group represent node by random select\\
	$\phi$ & total group number\\
    $\mathbb{PL}$ & package nodes list \\
    $\mathbb{PL}_k^{pa}$ & $k$th node in package list pack block success  \\
    $\mathbb{PL}_k^{ml}$ & $k$th node in package list pack block failed or initial attack \\
    $\mathbb{BC}_k$ & $k$th package node blockchain \\
    $\mathbb{BK}_k^{h1}$ & $k$th package node blockchain's block in height $h_1$ \\
    $\mathbb{MR}_k^{h1}$ & $k$th package node blockchain's block in height $h_1$ contains Merkle root\\
    \bottomrule
  \end{tabularx}
\end{table}

\begin{figure}[htbp]
    \centering
    \includegraphics[width=.8\textwidth]{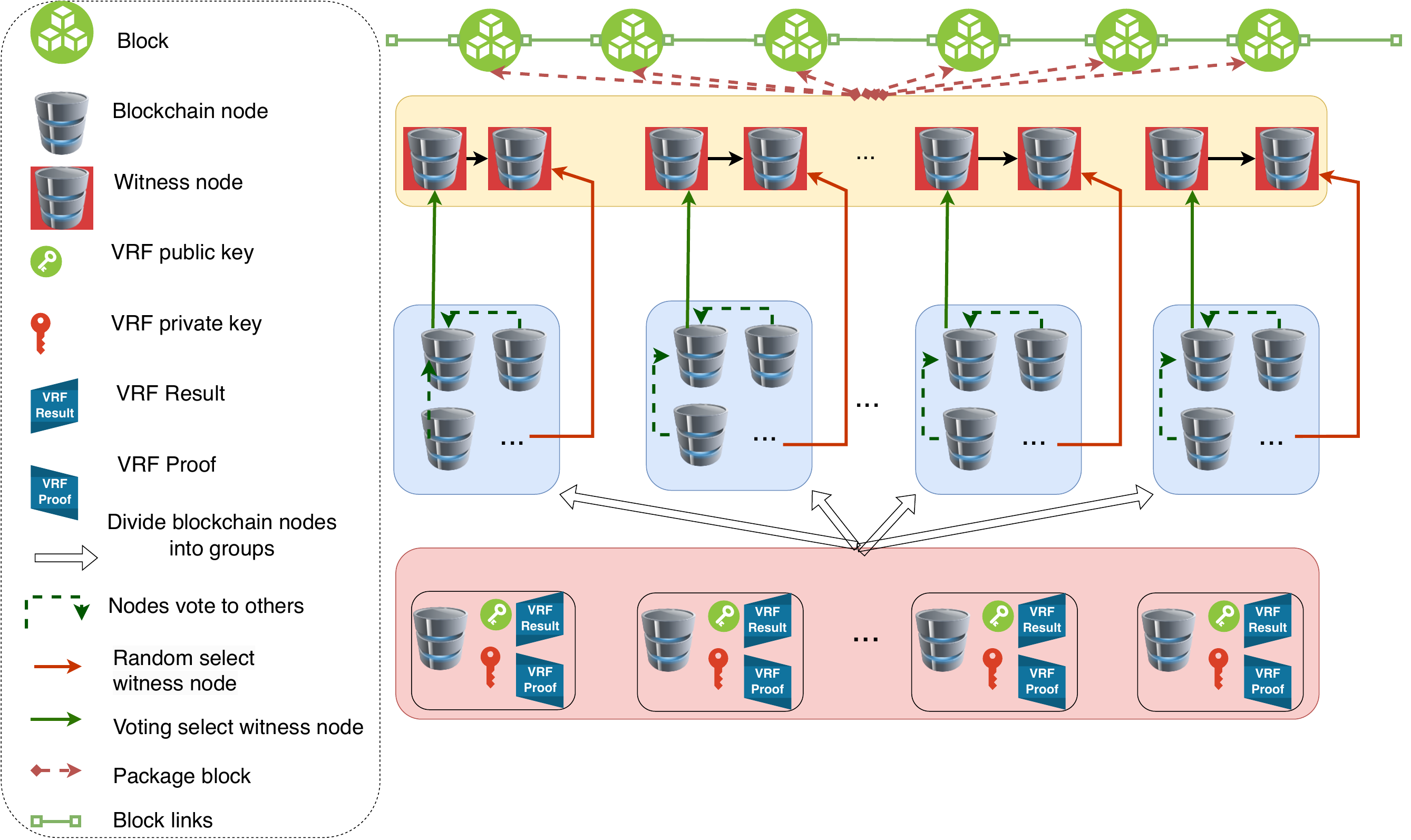}
    \caption{Schematic Diagram of HL-DPoS Consensus Algorithm}
    \label{Schematic of HL-DPoS Consensus Algorithm}
\end{figure} 

We improved the DPoS consensus algorithm by combining the consensus process with Game Theory, which proposed the HL-DPoS consensus algorithm, like the human society activity. Figure \ref{Schematic of HL-DPoS Consensus Algorithm} is the schematic diagram of HL-DPoS. We split the nodes into several groups by VRF. Every node votes for one of its group represent nodes and selects a node randomly by HL-DPoS algorithm as another represent node of their group to make nodes enter the witness nodes queue more randomly, reducing the centralization problem, which means every group has at least two nodes for witness nodes selects. All the groups represent nodes composed of the witness nodes queue, verifying blocks sequentially.  In HL-DPoS we grant witness nodes the right to supervise and report malicious behavior and can benefit from the reporting of malicious behavior. After the malicious attack node is deducted tokens, the deducted tokens will be rewarded to the reporting node. Once the attack is detected, there have several ways to punish attack nodes. The algorithm's main steps are shown in Algorithm \ref{alg:HL-DPoS Algorithm}.

\begin{algorithm}[H]
\caption{HL-DPoS Algorithm}
\label{alg:HL-DPoS Algorithm}
\SetAlgoLined
\SetKwInOut{Input}{Input}
\SetKwInOut{Output}{Output}

\SetKwFunction{generateVRFKeys}{generateVRFKeys}
\SetKwFunction{groupNodes}{groupNodes}
\SetKwFunction{selectBlockProducers}{selectBlockProducers}
\SetKwFunction{blockPackaging}{blockPackaging}
\SetKwFunction{rewardReporter}{rewardReporter}

\footnotesize

\Input{Blockchain nodes}
\Output{Block}


\SetKwProg{Main}{Main Function}{}{}
\Main{}{
    \tcp{HL-DPoS consensus algorithm}
    
    \groupNodes{}\;
    \selectBlockProducers{}\;
    \blockPackaging{}\;
    \rewardReporter{}\
}




\textbf{Function} \groupNodes{}: \quad \tcp{Node grouping}
\Indp
    $\psi \leftarrow 10$ \tcp{$\psi$ groups}
    
    
    \For{each node $N$}{
        ($privateKey$, $publicKey$) $\leftarrow$ \generateVRFKeys{}\;
        $proof \leftarrow$ computeVRFProof($privateKey$, $N$)\;
        $result \leftarrow$ computeVRFResult($proof$)\;
        $hashValue \leftarrow$ hash($result$)\;
        $groupIndex \leftarrow$ int($hashValue \times \psi$)\;
        addToGroup($groupIndex$, $N$)\;
    }
\Indm


\textbf{Function} \selectBlockProducers{}: \quad \tcp{Witness nodes selection}
\Indp
    $blockProducers \leftarrow []$\;
    
    
    \For{each group $G$}{
        $representatives \leftarrow []$\;
        
        
        \For{each node $N$ in group $G$}{
            $vote \leftarrow$ $N$.vote()\;
            $representatives$.append(($N$, $vote$))\;
        }
        
        
        $representative \leftarrow$ nodeWithMaxVotes($representatives$)\;
        $blockProducers$.append($representative$)\;
        
        
        $randomNode \leftarrow$ selectRandomNode($G$)\;
        $blockProducers$.append($randomNode$)\;
    }
    
    \KwRet $blockProducers$\;
\Indm


\textbf{Function} \blockPackaging{}: \quad \tcp{Block packaging operations}
\Indp
    \For{each block producer $P$}{
        $block \leftarrow P$.packageBlock()\;
        $reporter \leftarrow P$.reportAttack()\;
        
        
        \If{$report$ is not empty}{
            $fine \leftarrow$  HL-DPoS.deductTokens($P$, $fine$)\;
            \rewardReporter($reporter$, $fine$)\;
        }
    }
\Indm


\textbf{Function} \rewardReporter{}: \quad \tcp{Reward for reporting}
\Indp
    HL-DPoS.rewardTokens($reporter$, $fine$)\;
\Indm


\Indm\end{algorithm}

The algorithm work steps are as follows:
\begin{enumerate}
    \item {\bfseries\itshape VRF Generation.} All nodes in blockchain network $\mathbb{W}$ are grouped randomly according to VRF, first, we have to generate public key and private key for $VRF$. Randomly select a security parameter $\kappa = 2^{\iota}$, $\iota \in \mathbb{Z}^+$ . As $\iota$ increases, the security level also increases, but the time and space overhead for key generation and VRF computation also increase.  Select $a,b,p \in \mathbb{Z}$, makes elliptic curve $E_{p}(a,b):y^2\equiv x^3 + ax + b(\pmod{p})$, randomly choice a point $\mathbb{P}$ on $E_{p}(a,b)$ with  order $\mathbb{O}_\mathbb{P}$, where $E_{p}(a,b)$'s order $\mathbb{O}_{E_{p}(a,b)}$ is a large prime. Choice a security hash function $\mathbb{H} = hash(\mathbb{M})$, $\mathbb{M}$ is message to be hash, consider $\mathbb{H}$ length is $L_{\mathbb{H}}$, $\kappa$ length is $L_{\kappa}$, there should be  $L_{\kappa} \geq L_{\mathbb{H}}$. Randomly choose a value $\mathbb{V}$ from set interval $[1, \mathbb{O}_{E_{p}(a,b)}]$, calculate the point $\mathbb{PK}$ in $E_{p}(a,b)$ with function $\mathbb{PK} =\mathbb{V} \times \mathbb{P}$, which $\mathbb{PK}$ is the public key, and private key $\mathbb{SK} = (\kappa,\mathbb{V})$. Get a input $\mathbb{X}$, generate a random number $r$, do $\mathbb{SK}$ hash calculation by $\mathbb{H}$: $\mathbb{H}_h = hash(\mathbb{SK})$, concatenate $\mathbb{H}_h$ with message $\mathbb{X}$ to have the proof of VRF: $\mathbb{H}_c = hash(\mathbb{H}_h|\mathbb{X})$, the $|$ is concatenate action, the VRF calculate result is $s = r + \mathbb{H}_c \cdot \mathbb{V}$. Then we can get $D_l$'s VRF value with $\mathbb{X}$ : $\mathbb{VRF} = (\mathbb{H}_c,s)$. VRF of $D_l$ is $\mathbb{VRF}_{D_l} = (\mathbb{H}_c^{D_l},s^{D_l})$. 
    \item {\bfseries\itshape Group Cluster.} In this step, we split nodes in the blockchain network into several groups $\mathbb{G}_{D}$. For consensus security, we use the Hash function to transform $\mathbb{VRF}$ to a float number between set$[0,1]$. For $D_l$, the hash function being used to divid nodes is $\mathbb{H}_S(s^{D_l}) =  = \operatorname{SHA-256}(s^{D_l})$, $\mathbb{H}_S(s^{D_l}) \in [0,1]$.  Assume total group number is $\phi$,  if $\mathbb{H}_S(\mathbb{VRF}_{D_l}) \in [i-1/{\phi},i/{\phi}]$,$i\in [1,\phi]$, $D_l$ will be in the $i$th group, $\mathbb{G}_{D}^i$. The group cluster result can be verified by proof $\mathbb{H}_c^{D_l}$. For the next steps, if $\mathbb{G}_{D}$ only has one node, we will merge it with a random group, promise $\mathbb{G}_{D}$ contains more than two nodes.
    
  	In HL-DPoS, choose two different nodes from each group $\mathbb{G}_{D}$ to join the package nodes list $\mathbb{PL}$. For $\mathbb{G}_{D}^i$,  one of the two nodes, $R_v^i$, is selected by a group inside votes during the voting by the nodes in  $\mathbb{G}_{D}^i$. The other node, $R_s^i$, is randomly selected based on VRF. 
  	\item {\bfseries\itshape Voting Process.} $R_v^i$ is the voting node of $\mathbb{G}_{D}^i$. To choose which is $R_v^i$ from the nodes in $\mathbb{G}_{D}^i$, first, $D_l$ in $\mathbb{G}_{D}^i$ will vote to other nodes $D_{l_1}$, $l_1 \neq l$ and $l_1 \in \mathbb{G}_{D}^i$. Every node can vote to their trust mode. Each of them has one ticket, which they vote must be in the same group with them. Once the node they vote to be the $R_v^i$, they will share the benefits and the punishment for $R_v^i$'s package action. For the nodes who have tickets, how to find which one to vote for so that they can get more benefits instead of punishment? In DPoS consensus algorithm, the node with the highest stake will get the most tickets. In HL-DPoS, we still think nodes with higher stakes can package blocks in time and maintain the order of the blockchain network. But we set a few limitations on it. After the voting, sort the results from high to low, and get an ordered result list $\mathbb{RL}^i$. $\mathbb{G}_{D}^i$ contains nodes number is $\mathbb{NUM}_{\mathbb{G}_{D}^i}$, the $(i \bmod{(y-x)})$th node from the nodes in $\mathbb{RL}^i$ between $[\mathbb{NUM}_{\mathbb{G}_{D}^i} \times \sigma, \mathbb{NUM}_{\mathbb{G}_{D}^i} \times \tau]$ as $R_v^i$, $0 \leq x < y \leq 50\%$. Once a node voted to be $R_v^i$, in the next $\psi$ round it cannot become $R_v^i$ again. The $\psi$ is depend on $\mathbb{NUM}_{\mathbb{G}_{D}^i}$, $\psi = \mathbb{NUM}_{\mathbb{G}_{D}^i} \times \rho$, $\rho$ is a percentage number.  We split $D_l$ to several groups randomly, vote inside group, disable node join $\mathbb{PL}$ repeatedly within a certain range. 
 	\item {\bfseries\itshape Random Select Process.} We randomly select $R_s^i$ from $\mathbb{G}_{D}^i$, makes nodes join in $\mathbb{PL}$ have randomness. If a specific node enters the package queue multiple times in the consensus algorithm, the security of the blockchain network will be reduced, and the possibility of the specific node being attacked will be greater. First, for $\mathbb{G}_{D}^i$, exclude $R_v^i$ of this round and the previous $\psi$ round. Check if there have other types of nodes that cannot be $R_s^i$ and exclude them. Other types mean the past $\psi$ round's $R_s^i$, the nodes are forbidden to join $\mathbb{PL}$, etc. Second, take the last nodes into candidate queue $\mathbb{CD}^i$, the nodes number in $\mathbb{CD}^i$ is $\mathbb{NUM}_\mathbb{CD}^i$.   From $\mathbb{CD}^i$, the $\mathbb{CD}^i [\lfloor \text{rand()}\cdot n \rfloor]$ is the $R_s^i$, $rand() \in[0,1), seed \in \mathbb{R}$ is a random number, $n$ is $\mathbb{CD}^i$ contains nodes number.
  	\item {\bfseries\itshape Package Process.} $\mathbb{PL}$ contains $R_v^i$ and $R_s^i$ from $\mathbb{G}_{D}^i, i \in [1, \phi]$. Every group has two nodes included in $\mathbb{PL}$. We define the $k$th node in $\mathbb{PL}$ , $\mathbb{PL}_k, k \in[1, 2 \times \phi]$ has two possible actions, one is complete the block packaging work on time and accurately, the block it packs contains transactions in time, we think it’s a positive action, $\mathbb{PL}_k$ is a good node  $\mathbb{PL}_k^{pa}$. If $\mathbb{PL}_k$ didn't package blocks on time, or package wrong transactions in it, or modify the data on the blockchain, or the packaged block does not contain the transactions previously created by $\mathbb{PL}_k$, we will denote $\mathbb{PL}_k$ as a malicious node $\mathbb{PL}_k^{ml}$. Before nodes in $\mathbb{PL}$ prepared to be the witness node, they have to ensure that the tokens in their account $\mathbb{AT}$ are greater than the amount to be deducted, $PA$, by the consensus algorithm once  $\mathbb{PL}_k$ become the $\mathbb{PL}_k^{ml}$. If $\mathbb{PL}_k$ is a good node  $\mathbb{PL}_k^{pa}$, packs blocks in time and correctly, consensus will reward $\mathbb{PL}_k^{pa}$ with tokens $token_R$. If $\mathbb{PL}_k^{pa}$ is $R_v^i$, the nodes who vote for $\mathbb{PL}_k^{pa}$ in group $i$ will get reward with tokens $token_V$, $token_R \geq token_V$. If $\mathbb{PL}_k$ is $\mathbb{PL}_k^{ml}$, HL-DPoS will punish it in several ways.
  	\item {\bfseries\itshape Punishment Process.} \label{item:HL-punish} $\mathbb{PL}_k$ can verify the block packed by previous  $\mathbb{PL}_j, j<k$, check if $\mathbb{PL}_j$ did the right thing. Once if $\mathbb{PL}_j$ is a malicious node, $\mathbb{PL}_j^{ml}$, who  package blocks over time, package wrong transactions in the block, modify data on blockchain,  packaged block does not contain the transactions previously created by itself,  $\mathbb{PL}_k$ will has the power to report the malicious actions of $\mathbb{PL}_j^{ml}$ to HL-DPoS and receive token rewards after $\mathbb{PL}_j^{ml}$ malicious actions verified by HL-DPoS. The reward from malicious action can be considered as sheep, but once $\mathbb{PL}_j^{ml}$ eat the sheep, in HL-DPoS, $\mathbb{PL}_{k}$ will has the power to eat $\mathbb{PL}_j^{ml}$, which is the malicious action report.  The punishment of $\mathbb{PL}_j^{ml}$ include the following ways:
  		\begin{itemize}
  			\item Terminate $\mathbb{PL}_j^{ml}$ packaging rights, $\mathbb{PL}_{j+1}$ will take in charge of the package process if $\mathbb{PL}_{j+1}$  not from the same group with $\mathbb{PL}_j^{ml}$, otherwise $\mathbb{PL}_{j+2}$ will take in charge of the package process;
  			\item HL-DPoS deduct $\mathbb{PL}_j^{ml}$ $\mathbb{AT}$ with $PA$, for its malicious actions. If $\mathbb{PL}_l, j<l<k$ didn't discover $\mathbb{PL}_j^{ml}$ malicious action, $\mathbb{PL}_l$ will also be punish with $\mathbb{AT}$ deduct $PA$;
  			\item Kick $\mathbb{PL}_j^{ml}$ and the node in  $\mathbb{PL}$ from same group with $\mathbb{PL}_j^{ml}$ ,$\mathbb{PL}_{s}$,   out of $\mathbb{PL}$.
  			\item If $\mathbb{PL}_j^{ml}$ is a $R_v^i$, HL-DPoS will punish the nodes who voted to $\mathbb{PL}_j^{ml}$ in $i$th group, deduct their $\mathbb{AT}$ with $PV, PV < PA$.
  			\item The malicious action report node $\mathbb{PL}_k$ will get reward $reward_R$. $reward_R$ from the malicious nodes $\mathbb{AT}$ deduct,  $reward_R = PA \times (k-j)$ .
  			\end{itemize}
    \item {\bfseries\itshape Node Cycle.} After $\mathbb{PL}_k^{pa}$ packed block, it will not terminate this round package mission immediately, but join in the end of $\mathbb{PL}$, to verify $\mathbb{PL}_h, k< h \leq 2 \times \phi $, if it is a malicious node. After every package node in $\mathbb{PL}$ pack block done, HL-DPoS will enter next round. 

\end{enumerate}
%
%

\subsubsection{Longest Chain Verification Mechanism}

In HL-DPoS we keep $\mathbb{PL}$ contains both vote and random select nodes, give them the power to verify and report previous nodes package process, and reward the effect report. Once nodes do malicious action, the following nodes in $\mathbb{PL}$ can report the malicious nodes and get a reward from the malicious action report. If the following nodes don't report the malicious nodes, they will be reported by the nodes behind them. But even if we make random nodes enter $\mathbb{PL}$, prevent external attacks to specific nodes, report malicious actions, considering the worst-case scenario that there has one node in $\mathbb{PL}$, packaged a longer blockchain does not contain the transactions previously created by itself, and it colluded with other nodes in $\mathbb{PL}$, so no nodes reported the malicious action. Then it will generate a double-spend attack, which erases its previous transaction records. So for HL-DPoS, how to verify if the previous miner node excludes self previous proposed transactions? To address this issue, we proposed the longest chain verification mechanism based on the consensus blocks to solve the double-spend attack initiated by the $\mathbb{PL}$ node, which generates a new longest chain that doesn't contain transactions created by itself.  
\begin{figure}[htbp]
    \centering
    \includegraphics[width=.7\textwidth]{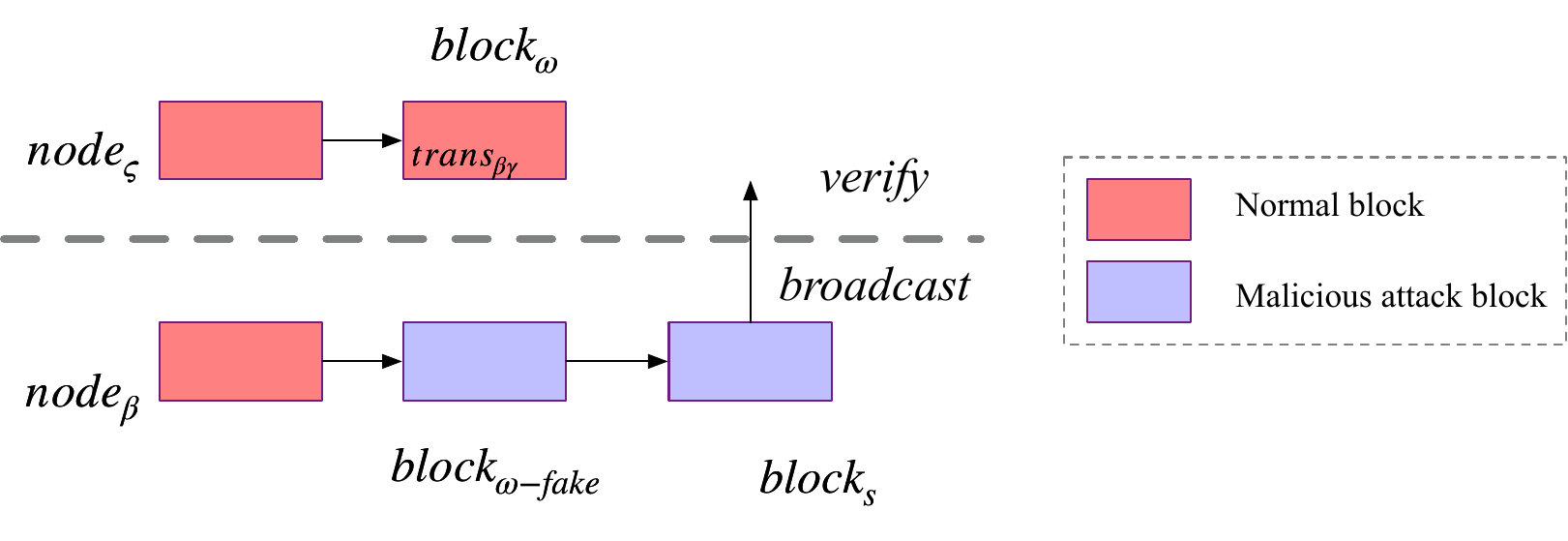}
    \caption{Schematic Diagram of Longest Chain Verification Mechanism}
    \label{Longest Chain Verification Mechanism}
    \end{figure} 

In HL-DPoS consensus algorithm, nodes will accept the longest chain as the main chain. To prevent the occurrence of $\mathbb{PL}$ nodes double-spend attack situation, after $\mathbb{PL}_k$ packed a block before it broadcasts its block, HL-DPoS will check the chain it broadcasts. Figure \ref{Longest Chain Verification Mechanism} shows the mechanism, and the mechanism work steps are as followings:
\begin{enumerate}[label=Step \arabic*]
	\item \label{item:first} $\mathbb{PL}_k$ broadcasts the block it generates, which means $\mathbb{PL}_k$ has the longest blockchain $\mathbb{BC}_k$. Check if there has same height block has a different hash id with other witness nodes $\mathbb{PL}_g, g \neq k \cap g \in [1, 2 \times \phi]$ accepted blockchain in $\mathbb{PL}$, $\mathbb{BC}_A$. If there has a block packed by $\mathbb{PL}_k$, $\mathbb{BK}_k^{h1}$, which height is $h1$, has a different hash id with the existing longest blockchain's same height block $\mathbb{BK}_g^{h1}, g \neq k$, which packed by $\mathbb{PL}_g$, then go to \ref{item:second}, else go to \ref{item:fourth};
	\item \label{item:second}Compare  $\mathbb{BK}_k^{h1}$'s Merkle root, $\mathbb{MR}_k^{h1}$ with $\mathbb{BK}_g$, check if the transactions set packed in  $\mathbb{BK}_k^{h1}$, $\mathbb{TR}_k^{h1}$, is same as transactions set in $\mathbb{BK}_g^{h1}$, $\mathbb{TR}_g^{h1}$. If $\mathbb{MR}_k^{h1} \neq \mathbb{MR}_g^{h1}$, which means $\mathbb{TR}_k^{h1} \neq \mathbb{TR}_g^{h1}$, then go to \ref{item:third}, else go to \ref{item:fourth};
	\item \label{item:third} Based on $\mathbb{TR}_g^{h1}$, compare transactions in $\mathbb{TR}_k^{h1}$ with $\mathbb{TR}_g^{h1}$, check if there has a transaction, $\mathbb{TR}_M$, contained in $\mathbb{TR}_g^{h1}$ but not in $\mathbb{TR}_k^{h1}$. If $\mathbb{TR}_k^{h1}$ has lost transactions that have already reached consensus, then check if $\mathbb{BK}^{h2}, h2 > h1$ contained $\mathbb{TR}_M$. If $\mathbb{BK}$ in $\mathbb{BC}$ miss $\mathbb{TR}_M$, and it is contained in  $\mathbb{BC}_A$, we will denote $\mathbb{PL}_k$ has hidden a transaction and attempted to broadcast the longest chain using its witness node identity to achieve long-range attack. Then it will turn to HL-DPoS consensus algorithm step \ref{item:HL-punish}, punish $\mathbb{PL}_k$, abandon $\mathbb{BC}_k$ and continue the block generation process with the next mining node. Else if there has no missing $\mathbb{TR}_M$, go to \ref{item:fourth}.
	\item  \label{item:fourth} Nodes in blockchain network accept $\mathbb{BC}_k$, as the mainchain which reached consensus.
\end{enumerate}
	
In Algorithm \ref{alg:verify-longest-chain} shows how the longest chain verification mechanism works, we assumed $node_\beta$ is a malicious node and a witness node. It initialed a transaction of tokens from itself to $node_{\gamma}$ in the previous block and mined a new block to exclude its transaction. While it takes turns to package the block, it adds blocks after the blocks it mined so that the transactions it initiated earlier are excluded from the longest chain, which means the record of $trans_{\beta \gamma}$ is erased. $node_\beta$ still has the tokens it transferred to $node_{\gamma}$  before. In this case, our proposed longest-chain verification mechanism will compare the transactions in the longest chain and the consensuses chain in other nodes to check if $trans_{\beta \gamma}$ is missed. If the longest chain has missed transactions in the consensuses chain, the longest chain will not be accepted.

\begin{algorithm}[H]
\SetAlgoLined
\SetKwInOut{Input}{Input}
\SetKwInOut{Output}{Output}
\caption{Verify Longest Chain}
\label{alg:verify-longest-chain}

\Input{ReceivedChain (the longest chain received)}
\Output{Acceptance of the longest chain as the main chain}

\BlankLine

$transactionsExist \gets \text{True}$;

\ForEach{block in node's Blockchain}{
\ForEach{transaction  in block}{
\If{transaction not in ReceivedChain}{
$transactionsExist \gets \text{False}$;
\textbf{break};
}
}}

\If{$transactionsExist$}{
\Return \text{True}. \tcp{Accept the longest chain as the main chain}
}
\Else{
\Return \text{False}. \tcp{Do not accept the longest chain as the main chain}
}

\end{algorithm}

\section{Algorithm Analysis}
\label{Algorithm Analysis}





\subsection{HL-DPoS security analysis}
\label{HL-DPoS security analysis}
During the packaging process, based on DEFINITION \ref{HL-DPoS Consensus Algorithm}, assume nodes in blockchain network number is $\mathbb{NUM}_{\mathbb{D}}$, witness nodes queue contains nodes number is  $\mathbb{NUM}_{\mathbb{PL}} = \varkappa, \varkappa \in \mathbb{Z}^+$ , $ \delta$ can be represent as Equation \ref{D_l}, \ref{A_i}, \ref{T}, \ref{S}, \ref{P}, \ref{I}, \ref{u}. Each node can supervise whether the previous nodes have behaved maliciously or covered up the behavior of previous malicious nodes. 

In this case, we define a node that performs normally can get profit $P_1$, and perform maliciously can get profit $P_2$ but may be punished by other nodes. If being punished, then the profit is $-\varrho \times P_2, \varrho \in \mathbb{Z}^+$, if the following node didn't prevent previous malicious nodes, it will be considered malicious, who punish the malicious node can get every malicious nodes' fine.

\begin{equation}
D_l=\{D_1,D_2,\ldots,D_{\mathbb{NUM}_{\mathbb{D}}}\}
\label{D_l}
\end{equation}

In Equation \ref{D_l}, $D_l$ has $\mathbb{NUM}_{\mathbb{D}}$ samples, from $D_1$ to $D_{\mathbb{NUM}_{\mathbb{D}}}$, represent different nodes.

\begin{equation}
	\mathbb{PL}_{\varepsilon} = \{ \mathbb{PL}_1, \mathbb{PL}_2, \ldots, \mathbb{PL}_{\varkappa} \}
	\label{PL}
\end{equation}
 
In Equation \ref{PL}, witness nodes queue has $\varkappa$ nodes, from $\mathbb{PL}_1$ to $\mathbb{PL}_{\varkappa}$, and $\mathbb{PL}_{\varepsilon} \in D_l$.

\begin{equation} 
A_i=\{A_1,A_2,\ldots,A_{\varkappa}\}
\label{A_i}
\end{equation}

In Equation \ref{A_i}, there has $\varkappa$ action set, for witness nodes $\mathbb{PL}_1$ to $\mathbb{PL}_{\varkappa}$, each of them has one action set, every action set is a collection of actions that the corresponding nodes may make. In this case, the action set includes both malicious behavior (initiating attacks or harboring malicious nodes) and normal behavior (normal packaging without harboring malicious nodes).

\begin{equation}
\begin{aligned}
T=\{(N_{\mathbb{PL}_1}),(M_{\mathbb{PL}_1},N_{\mathbb{PL}_2}),(M_{\mathbb{PL}_1},M_{\mathbb{PL}_2},N_{\mathbb{PL}_3}),\\
(M_{\mathbb{PL}_1},M_{\mathbb{PL}_2},M_{\mathbb{PL}_3},\ldots, M_{\mathbb{PL}_{\varpi}}, \ldots, M_{\mathbb{PL}_{\varkappa - 1}}, N_{\mathbb{PL}_{\varkappa}}),\\
(M_{\mathbb{PL}_1},M_{\mathbb{PL}_2},M_{\mathbb{PL}_3},\ldots,  M_{\mathbb{PL}_{\varkappa}})
\end{aligned}
\label{T}
\end{equation}

In Equation \ref{T} has all terminal histories of $\mathbb{PL}_1$ to $\mathbb{PL}_{\varkappa}$, in this case T contains $(N_{\mathbb{PL}_1})$ indicates $\mathbb{PL}_1$ is a normal node, $(M_{\mathbb{PL}_1},N_{\mathbb{PL}_2})$ means $\mathbb{PL}_1$ is a malicious node, $\mathbb{PL}_2$ is a normal node which didn't harbor $\mathbb{PL}_1$, $(M_{\mathbb{PL}_1},M_{\mathbb{PL}_2},N_{\mathbb{PL}_3})$ means $\mathbb{PL}_1$,$\mathbb{PL}_2$ are malicious nodes, $\mathbb{PL}_3$ is a normal node which didn't harbor $\mathbb{PL}_1$ and $\mathbb{PL}_2$, $(M_{\mathbb{PL}_1}$,$M_{\mathbb{PL}_2}$, $M_{\mathbb{PL}_3}$, $\ldots$, $M_{\mathbb{PL}_{\varpi}}$, $\ldots$,$ M_{\mathbb{PL}_{\varkappa - 1}}$, $N_{\mathbb{PL}_{\varkappa}})$ means for $\mathbb{PL}$ nodes beside the last one $\mathbb{PL}_{\varkappa}$ are malicious nodes, $\mathbb{PL}_{\varkappa}$ is a normal node which didn't harbor previous malicious nodes. $(M_{\mathbb{PL}_1},M_{\mathbb{PL}_2},M_{\mathbb{PL}_3},\ldots,  M_{\mathbb{PL}_{\varkappa}})$ means every nodes in $\mathbb{PL}$ are malicious nodes.
 
\begin{equation}
S_T=
\{
\begin{aligned}\eta,N_{\mathbb{PL}_1},M_{\mathbb{PL}_1},N_{\mathbb{PL}_2},M_{\mathbb{PL}_2},N_{\mathbb{PL}_3},\\
M_{\mathbb{PL}_3}, \ldots, N_{\mathbb{PL}_{\varpi}}, M_{\mathbb{PL}_{\varpi}}, \ldots, N_{\mathbb{PL}_{\varkappa}},M_{\mathbb{PL}_{\varkappa}},
\end{aligned}\}
\label{S}
\end{equation}

Equation \ref{S} has all terminal histories‘ proper sub-histories, $\eta$ means empty, $N$ means the node behaves normally, it will package blocks in time and prevent the possible malicious attack from previous nodes. $M$ means the node is a malicious node, it will initiate an attack while package blocks or don't package blocks in time.

\begin{equation}
\begin{aligned}
P(\eta)=\mathbb{PL}_1,P(N_{\mathbb{PL}_1})=\mathbb{PL}_1,P(M_{\mathbb{PL}_1})=\mathbb{PL}_1,\\
P(N_{\mathbb{PL}_2})=\mathbb{PL}_2, P(M_{\mathbb{PL}_2})=\mathbb{PL}_2,\\
P(N_{\mathbb{PL}_3})=\mathbb{PL}_3, P(M_{\mathbb{PL}_3})=\mathbb{PL}_3,\\
\ldots \\
P(N_{\mathbb{PL}_{\varpi}})=\mathbb{PL}_{\varpi},P(M_{\mathbb{PL}_{\varpi}})=\mathbb{PL}_{\varpi},\\
\ldots \\
P(N_{\mathbb{PL}_{\varkappa}})=D_{\varkappa}, P(M_{\mathbb{PL}_{\varkappa}})=\mathbb{PL}_{\varkappa}
\end{aligned}
\label{P}
\end{equation}

In Equation \ref{P} denotes sub-history and corresponding participating node, $\eta$, $N_{\mathbb{PL}_1}$ and $M_{\mathbb{PL}_1}$ participating node is $\mathbb{PL}_1$. $N_{\mathbb{PL}_2}$, $M_{\mathbb{PL}_2}$ participating node is $\mathbb{PL}_2$. $N_{\mathbb{PL}_3}$, $M_{\mathbb{PL}_3}$ participating node is $\mathbb{PL}_3$. For nodes $\mathbb{PL}_{\varpi}, 3 < \varpi < \varkappa $,  $N_{\mathbb{PL}_{\varpi}}$ and $M_{\mathbb{PL}_{\varpi}}$ participating node is $\mathbb{PL}_{\varpi}$. The last node in $\mathbb{PL}$, $\mathbb{PL}_{\varkappa}$, is the participate node of $N_{\mathbb{PL}_{\varkappa}}$ and $M_{\mathbb{PL}_{\varkappa}}$.

\begin{equation}
\begin{aligned}
I_1=\{\{N_{\mathbb{PL}_1}\},\{M_{\mathbb{PL}_1}\}\},I_2=\{\{N_{\mathbb{PL}_2}\},\{M_{\mathbb{PL}_2}\}\},\\
I_3=\{\{N_{\mathbb{PL}_3}\},\{M_{\mathbb{PL}_3}\}\}, \ldots, I_{\varpi}=\{\{N_{\mathbb{PL}_{\varpi}}\},\{M_{\mathbb{PL}_{\varpi}}\}\},\\
\ldots, I_{\varkappa}=\{\{N_{\mathbb{PL}_{\varkappa}}\},\{M_{\mathbb{PL}_{\varkappa}}\}\},
\end{aligned}
\label{I}
\end{equation}

In Equation \ref{I} every $I$ contains corresponding nodes' information sets, $I_1$ contains $\mathbb{PL}_1$ two information sets, $N_{\mathbb{PL}_1}$ is the normal information set, $M_{\mathbb{PL}_1}$ is the malicious information set. Same as $I_2$ contains $\mathbb{PL}_2$ two information sets, $I_3$ contains $\mathbb{PL}_3$ two information sets,  $I_{\varpi}, 3 < \varpi < \varkappa$ contains $D_{\varpi}$ two information sets,  and  $I_{\varkappa}$ contains $\mathbb{PL}_{\varkappa}$ two information sets.

\begin{equation}
\begin{aligned}
\left\{
        \begin{array}{ll}
u_1(N_{\mathbb{PL}_1})=P_1,\\ 
u_1(M_{\mathbb{PL}_1} N_{\mathbb{PL}_2})=-\varrho \times P_2,\\ 
u_1(M_{\mathbb{PL}_1} M_{\mathbb{PL}_2} N_{\mathbb{PL}_3})=-\varrho \times P_2,\\
u_1(M_{\mathbb{PL}_1} M_{\mathbb{PL}_2} M_{\mathbb{PL}_3} M_{\mathbb{PL}_{\ldots}} N_{\mathbb{PL}_{\varkappa}})=-\varrho \times P_2, \\
u_1(M_{\mathbb{PL}_1} M_{\mathbb{PL}_2} M_{\mathbb{PL}_3} M_{\mathbb{PL}_{\ldots}} M_{\mathbb{PL}_{\varkappa}})=P_2, \\
\ldots \\
u_{\varpi}(N_{\mathbb{PL}_1}N_{\mathbb{PL}_2}N_{\mathbb{PL}_{\ldots}}N_{\mathbb{PL}_{\varpi}})=P_1,\\
u_{\varpi}(N_{\mathbb{PL}_1}N_{\mathbb{PL}_2}N_{\mathbb{PL}_{\ldots}}N_{\mathbb{PL}_{\varpi-1}}M_{\mathbb{PL}_{\varpi}}M_{\mathbb{PL}_{\varpi+1}}\\
M_{\mathbb{PL}_{\ldots}}M_{\mathbb{PL}_{\varkappa-1}}N_{\mathbb{PL}_{\varkappa}})\\
=-\varrho \times P_2,\\
u_{\varpi}(M_{\mathbb{PL}_1}M_{\mathbb{PL}_2}M_{\mathbb{PL}_{\ldots}}M_{\mathbb{PL}_{\varpi-1}}M_{\mathbb{PL}_{\varpi}}M_{\mathbb{PL}_{\varpi+1}}\\
M_{\mathbb{PL}_{\ldots}}M_{\mathbb{PL}_{\varkappa-1}}M_{\mathbb{PL}_{\varkappa}})\\
=P_2,\\
\ldots \\
u_{\varkappa}(N_{\mathbb{PL}_1}N_{\mathbb{PL}_2}N_{\mathbb{PL}_{\ldots}}N_{\mathbb{PL}_{\varkappa}})=P_1,\\
u_{\varkappa}(N_{\mathbb{PL}_1}N_{\mathbb{PL}_2}N_{\mathbb{PL}_{\ldots}}N_{\mathbb{PL}_{\varpi-1}}M_{\mathbb{PL}_{\varpi}}M_{\mathbb{PL}_{\varpi+1}}\\
M_{\mathbb{PL}_{\ldots}}M_{\mathbb{PL}_{\varkappa-1}}N_{\mathbb{PL}_{\varkappa}})\\
=\varrho \times P_2 \times (\varkappa - \varpi) + P_1,\\
u_{\varkappa}(N_{\mathbb{PL}_2}N_{\mathbb{PL}_{\ldots}}N_{\mathbb{PL}_{\varpi-1}}M_{\mathbb{PL}_{\varpi}}M_{\mathbb{PL}_{\varpi+1}}M_{\mathbb{PL}_{\ldots}}\\
M_{\mathbb{PL}_{\varkappa-1}}M_{\mathbb{PL}_{\varkappa}}N_{\mathbb{PL}_1})\\
=-\varrho \times P_2,\\
u_{\varkappa}(M_{\mathbb{PL}_1}M_{\mathbb{PL}_2}M_{\mathbb{PL}_{\ldots}}M_{\mathbb{PL}_{\varpi-1}}M_{\mathbb{PL}_{\varpi}}M_{\mathbb{PL}_{\varpi+1}}\\
M_{\mathbb{PL}_{\ldots}}M_{\mathbb{PL}_{\varkappa-1}}M_{\mathbb{PL}_{\varkappa}}M_{\mathbb{PL}_1})\\
=P_2,\\
        \end{array}
    \right.
\end{aligned}
\label{u}
\end{equation}

Equation \ref{u} denotes every nodes' utility, $u_1(N_{\mathbb{PL}_1})=P_1$ denote while $\mathbb{PL}_1$ is a normal node, its profit is $P_1$, so $u_1 =P_1$. The $k$ in $u$ denotes different nodes id, $M$ and $N$ in Equation \ref{u} denotes whether the behavior of the node corresponding to the serial number is malicious or normal. So $u_1(M_{\mathbb{PL}_1} N_{\mathbb{PL}_2})=-\varrho \times P_2$, denotes $\mathbb{PL}_1$ is a malicious node, $\mathbb{PL}_2$ is a normal node, $\mathbb{PL}_2$ prevent and punished $\mathbb{PL}_1$ malicious behavior, so $\mathbb{PL}_1$'s profit is $-\varrho \times P_2$. etc. $u_{\varpi}(N_{\mathbb{PL}_1}N_{\mathbb{PL}_2}N_{\mathbb{PL}_{\ldots}}N_{\mathbb{PL}_{\varpi-1}}M_{\mathbb{PL}_{\varpi}}M_{\mathbb{PL}_{\varpi+1}}M_{\mathbb{PL}_{\ldots}}M_{\mathbb{PL}_{\varkappa-1}}\\N_{\mathbb{PL}_{\varkappa}}) =-\varrho \times P_2 $ means for node $\mathbb{PL}_{\varpi}, 1 < \varpi < \varkappa$, if it is a malicious node and other nodes report the malicious action, it will be punished, and the profit will be $-\varrho \times P_2$. But if all the $\mathbb{PL}$ nodes are malicious nodes, their profit will be $P_2$. The last node always has a positive profit. If previous nodes are normal and it also performs normal, it can get $P_1$. If previous nodes have malicious nodes, such as from $\mathbb{PL}_{\varpi}$ to $\mathbb{PL}_{\varkappa -1}$, and it is a normal node, $\mathbb{PL}_{\varkappa}$ will report previous malicious nodes, and get profit for both normal action and report action, the profit will be $\varrho \times P_2 \times (\varkappa - \varpi)$. $\mathbb{PL}$ nodes will push to the end after the packed block to watch the behind nodes, so while every node in $\mathbb{PL}$ is malicious, $u_{\varkappa}=P_2$. Based on the Nash Equilibrium strategy, every node will choose a strategy that makes its profit the greatest. So in Equation \ref{u}, if $\mathbb{PL}_{\varkappa}$ choose to be a normal node, it may get profit $\varrho \times P_2 \times (\varkappa - \varpi) + P_1$ or $P_1$. But if it is a malicious node, it may get profit $-\varrho \times P_2$ or $P_2$. While $\mathbb{PL}_{\varkappa}$ is a normal node, its profit is always positive, but when $\mathbb{PL}_{\varkappa}$ is a malicious node, it may be punished. So nodes in $\mathbb{PL}$ behave normally is a Nash equilibrium point. It's the best choice for $\mathbb{PL}$ nodes. From Equation \ref{u}, we can find that the fault tolerance of HL-DPoS package queue is $(\mathbb{NUM}_{\mathbb{PL}}-1)/\mathbb{NUM}_{\mathbb{PL}}$.

\subsection{Long Range Attack Resistance Analysis}
\label{Long Range Attack Resistance Analysis}
We now analyze the probability $Pb$ of nodes in $\mathbb{PL}$ behave normally or maliciously. We use $Pb_{\varpi}^{N}$ represents node $\mathbb{PL}_{\varpi}$ behaves normal, $Pb_{\varpi}^{N} \in [0,1] \cap \mathbb{R}$, $Pb_{\varpi}^{N} = 0$ represents $\mathbb{PL}_{\varpi}$ is a malicious node, and $Pb_{\varpi}^{N} = 1$ represents $\mathbb{PL}_{\varpi}$ is a normal node. $Pb_{\varpi}^{M} = 1-Pb_{\varpi}^{N}$ represents $\mathbb{PL}_{\varpi}$ initial a long-range attack, create a new chain tried to manipulate transactions record in the blockchain. Denote $E_{\varpi}$ as $\mathbb{PL}_{\varpi}$ average profit from package process. From section \ref{HL-DPoS security analysis} we can find out that total $\mathbb{PL}$ nodes number is $\varkappa$, pack block normally can get profit $P_1$, behave maliciously, and if being reported then profit is $-\varrho \times P_2$, if $\mathbb{PL}$ nodes report other malicious nodes then the profit will get all the fine as profit. So $\mathbb{PL}_{\varpi}$ average profit is:

\begin{equation}
\begin{aligned}
E_{\mathbb{PL}_{\varpi}}=&Pb_{1}^{N} \times \ldots \times Pb_{\varpi}^{N} \times P_1\\
&+Pb_{1}^{N} \times \ldots \times (1-Pb_{\varpi - {\lambda_1} }^{N}) \times \ldots \\
&\times  Pb_{\varpi}^{N} \times ( {\lambda_1} \times \varrho \times P_2 + P_1)\\
&+(1-Pb_{\varpi}^{N}) \times \ldots \times Pb_{\varpi+\lambda_2}^{N} \times (-\varrho \times P_2)\\
&+(1-Pb_{1}^{N}) \times \ldots \times (1-Pb_{\varpi}^{N}) \times \ldots \\
&\times (1-Pb_{\varkappa}^{N}) \times P_2\\
\end{aligned}
\label{ED1}
\end{equation}

Hence, $0 < \lambda_1,\lambda_2 < \varpi < \varkappa$. Assume $\mathbb{PL}_{\varpi}$ is a malicious node, if the probability $Pb_{\varpi+1}^{N} $ of $\mathbb{PL}_{\varpi+1}$ is 0, which means $\mathbb{PL}_{\varpi+1}$ is a malicious node, when $\mathbb{PL}_{\varpi+1}$ fails to stop and punish $\mathbb{PL}_{\varpi}$ malicious behavior in a timely manner, the expected benefit obtained by $\mathbb{PL}_{\varpi+1}$ is:

 \begin{equation}
\begin{aligned}
 E_{\mathbb{PL}_{\varpi+1}^M}=
 \\ (1-Pb_{\varpi+1}^{N}) \times \ldots \times (1 - Pb_{\varpi+\lambda_2-1}^{N})
 \\ \times Pb_{\varpi+\lambda_2}^{N}  \times (-\varrho \times P_2) 
 \\ + (1-Pb_{\varpi+1}^{N}) \times \ldots \times (1-Pb_{\varkappa}^{N}) \times P_2\\
\end{aligned}
 \end{equation}

If $Pb_{\varpi+1}^{N}$ is 1, that is, when $\mathbb{PL}_{\varpi+1}$  detects malicious behavior of $\mathbb{PL}_{\varpi}$  and punishes it, the expected profits obtained by $\mathbb{PL}_{\varpi+1}$ is:

 \begin{equation}
 E_{\mathbb{PL}_{\varpi+1}^N}=\lambda_1  \times \varrho \times P_2 + P_1
 \end{equation}

 Hence, our HL-DPoS consensus algorithm is secure from the long-range attack when choosing proper $P_1$ and $\varrho$, makes $ E_{\mathbb{PL}_{\varpi+1}^N} \gg E_{\mathbb{PL}_{\varpi+1}^M}$.

For example, assume $\lambda_1 =1, \lambda_2 = 1$, $\mathbb{PL}_{\varpi+1}$ is a malicious node, and consider the extreme situation that $\mathbb{PL}_{\varpi+1}$ can get greatest benefit, from $\mathbb{PL}_{\varpi+1}$ to $\mathbb{PL}_{\varkappa}$ are malicious nodes. So $(1-Pb_{\varpi+1}^{N}) = 1$, $(1-Pb_{\varkappa}^{N}) = 1$, $E_{\mathbb{PL}_{\varpi+1}^M}= 1 \times 1 \times \ldots \times P_2 = P_2$. And if $\mathbb{PL}_{\varpi+1}$ is a normal node, consider the extreme situation that $\mathbb{PL}_{\varpi+1}$ can get lowest benefit, that is, no malicious node before $\mathbb{PL}_{\varpi+1}$, so $E_{\mathbb{PL}_{\varpi+1}^N} = P_1$. So one of the feasible solution to make $ E_{\mathbb{PL}_{\varpi+1}^N} \gg E_{\mathbb{PL}_{\varpi+1}^M}$ is $P_1 \gg P_2$. Once there has one normal node that detect the $\mathbb{PL}_{\varpi+1}$ is a malicious node, the $E_{\mathbb{PL}_{\varpi+1}^M}$ will be $-\varrho \times P_2 < 0$, so even if $P_1 \leq P_2$, we can also set $\varrho$ to a large number to warning malicious nodes, to reduce the probability of long-range attack.

\section{SIMULATION AND RESULT ANALYSIS}
In this section, we used blockchain simulator \cite{Alharby2020BlockSimAE} to compare the experimental results of the PoW , DPoS , and our proposed HL-DPoS consensus algorithms. The simulation program runs on a MacBook Air with an M1 chip, 16GB of memory, and a 512GB hard drive. We set the number of DPoS witness nodes to 101, HL-DPoS witness nodes to 120, which is for the nodes in group split can be divisible, and in HL-DPoS all nodes are divided into 10 groups. We mainly analyze the performance of HL-DPoS Algorithm in two fields: the number of blocks changes with different number of nodes; the number of validating nodes changes over different periods of time.

\subsection{SIMULATION DESIGN}
In HL-DPoS we split nodes into several groups while the nodes voting section selects witness nodes in two different ways, using malicious nodes report and longest chain verify mechanism to prevent witness nodes conduct malicious behavior or launching a long-range attack. 

{\bfseries\itshape PoW implementation} The PoW consensus algorithm being implemented in blockchain simulator Bitcoin, according to model the propagation of transactions and blocks as time delay to abstract the underlying broadcast protocol, based on Maher Alharby \cite{Alharby2020BlockSimAE}, the source data from DSN Bitcoin monitoring to parameterize the model, from where to obtain the transmission delay of information, the block interval being set to 600 seconds.

{\bfseries\itshape DPoS implementation} DPoS implement in blockchain simulator Ethereum, the propagation of transactions and blocks as a time delay to abstract the Ethereum underlying broadcast protocol base, based on Maher Alharby\cite{Alharby2020BlockSimAE},  the block interval being set to 12.42 seconds. We implement the voting process. Nodes can vote to elect their trust nodes into package nodes list being witness nodes, and witness nodes can package blocks in turn.

{\bfseries\itshape Transaction Implementation } Transactions are modeled by a single transaction pool shared among all nodes in the network, from Maher Alharby\cite{Alharby2020BlockSimAE}.


\begin{figure}[htbp]
	\centering
	\subfloat[10 minutes]{\includegraphics[width=.45\columnwidth]{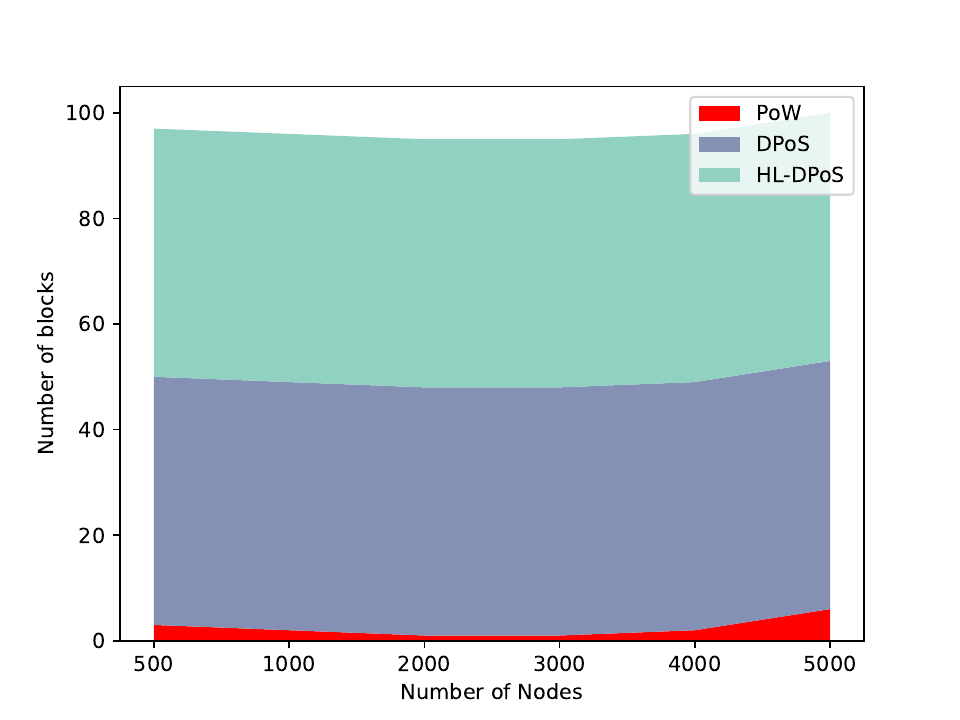}}\hspace{5pt}
        \subfloat[20 minutes]{\includegraphics[width=.45\columnwidth]{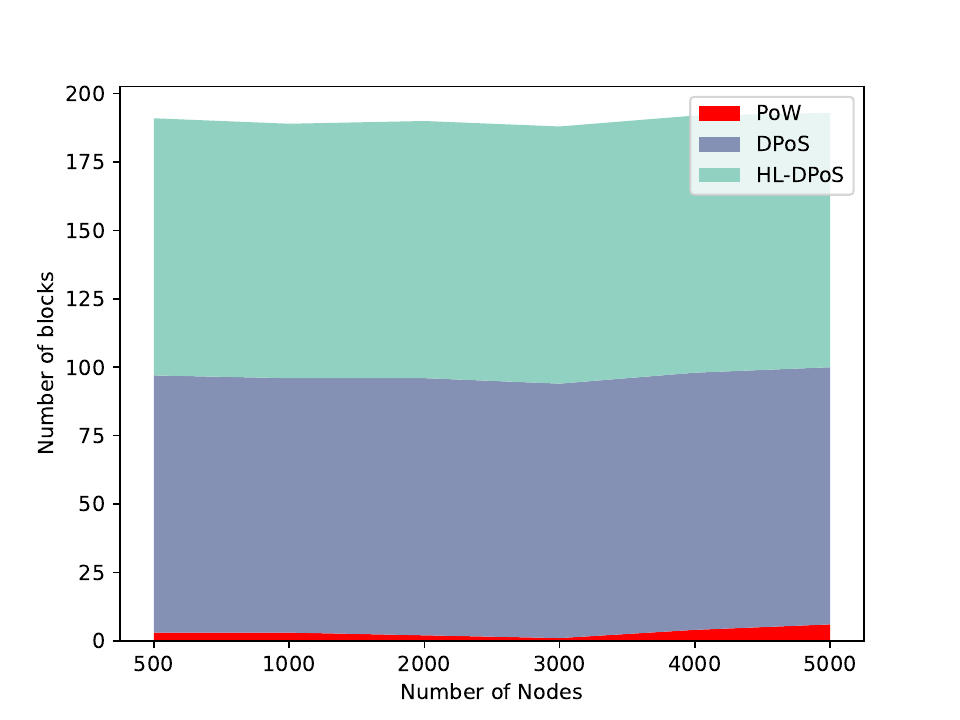}}\\
        
        \subfloat[30 minutes]{\includegraphics[width=.45\columnwidth]{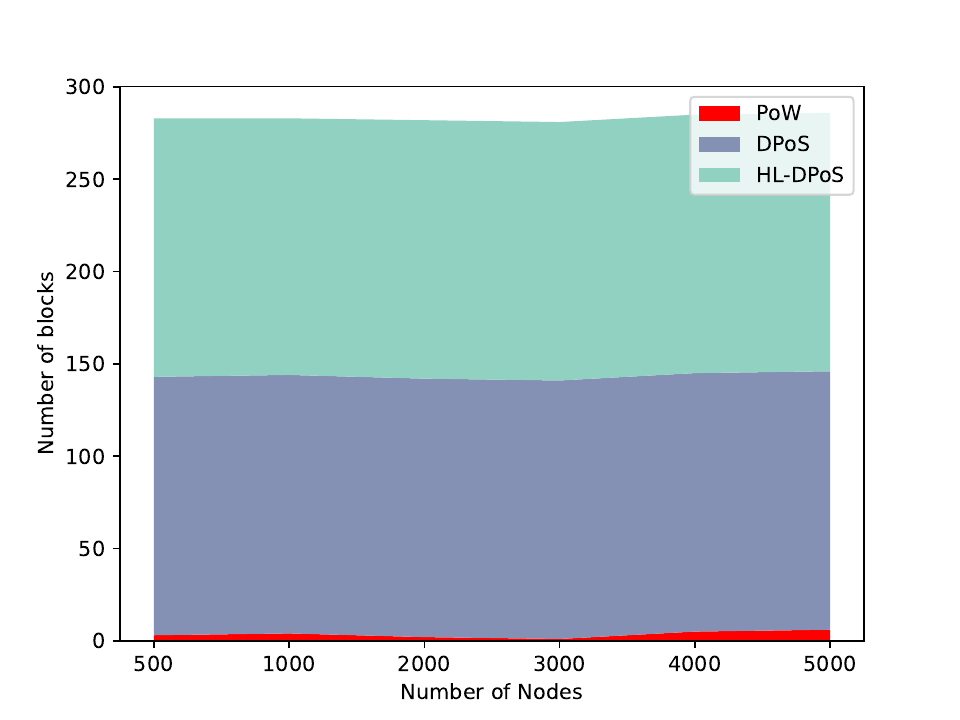}}\hspace{5pt}
	\subfloat[40 minutes]{\includegraphics[width=.45\columnwidth]{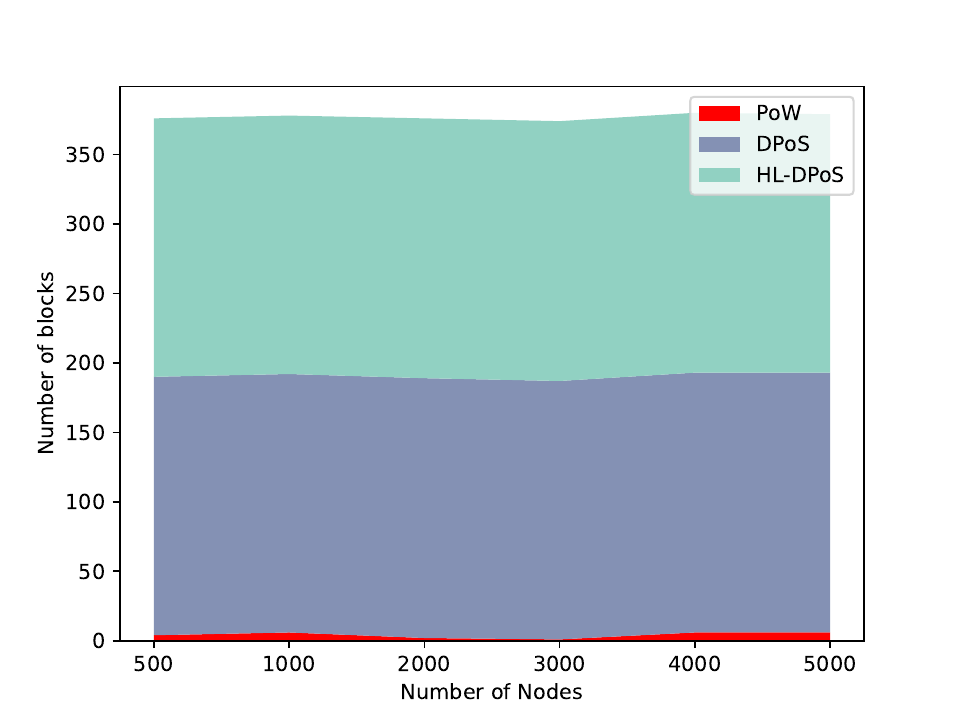}}\\
	
        \subfloat[50 minutes]{\includegraphics[width=.45\columnwidth]{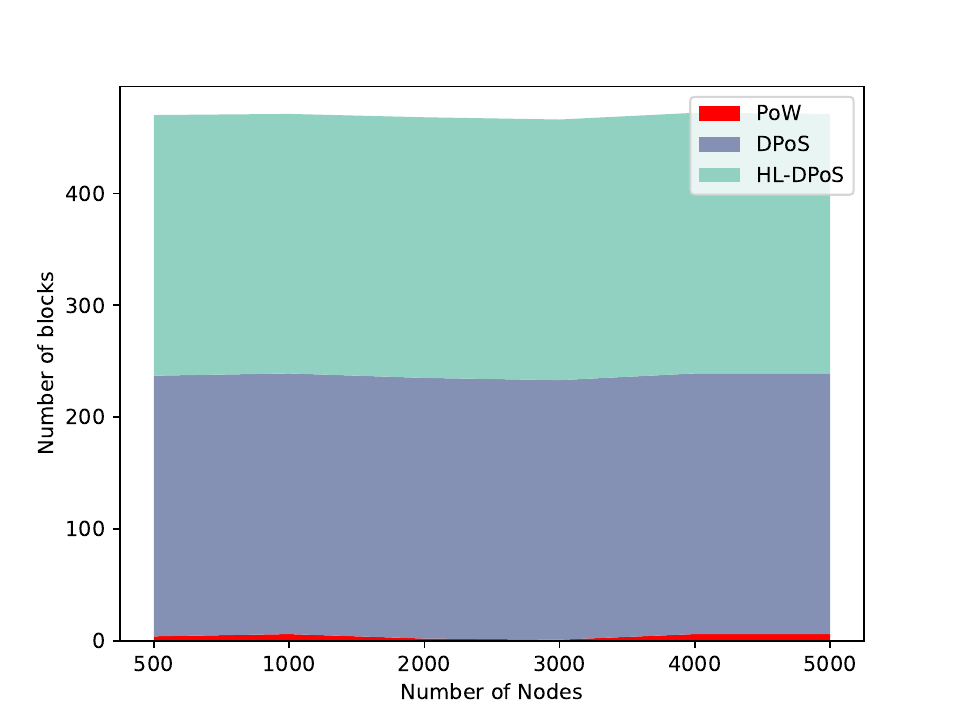}}\hspace{5pt}
        \subfloat[60 minutes]{\includegraphics[width=.45\columnwidth]{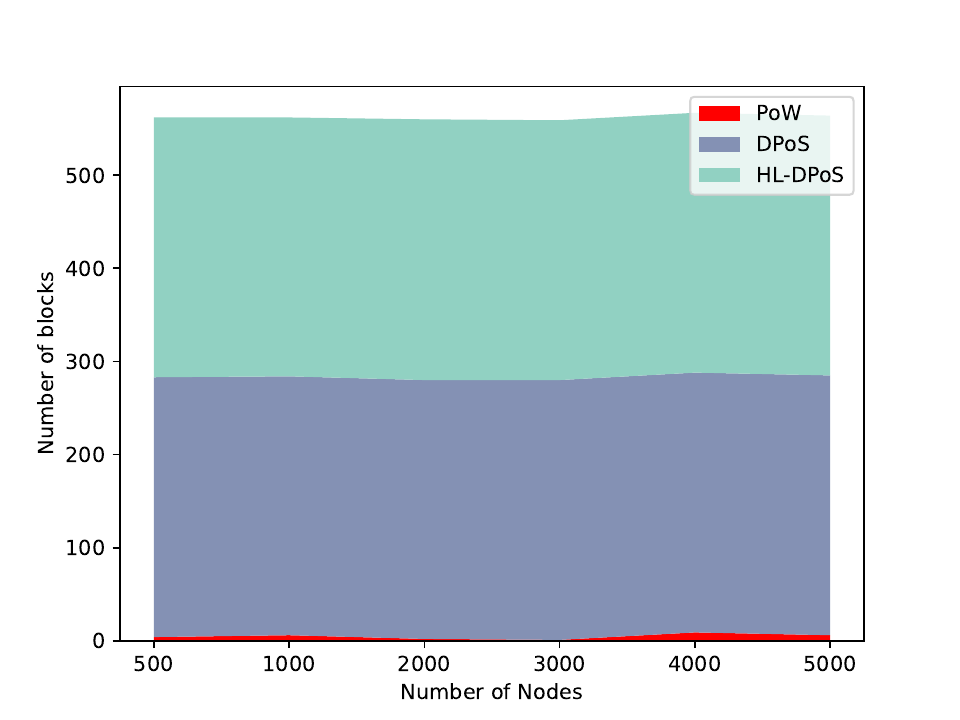}}\\
	\caption{ Number of Blocks Generates with the Number of Nodes (a) 10 minutes. (b) 20 minutes. (c) 30 minutes. (d) 40 minutes. (e) 50 minutes. (f) 60 minutes. }
\label{Number of Blocks Changes with the Number of Nodes.}   
\end{figure}

We want to prove HL-DPoS has high efficiency like DPoS with high security, and the security has been analyzed in Section \ref{Algorithm Analysis}. Figure \ref{Number of Blocks Changes with the Number of Nodes.} shows HL-DPoS has a high level of efficiency compared with DPoS, greater than PoW, Figure \ref{Times of generating blocks} shows HL-DPoS has fewer witness nodes entering package queue repeatedly, which means HL-DPoS has less centralization problem in HL-DPoS.

\subsection{Throughput emulation}
 Figure \ref{Number of Blocks Changes with the Number of Nodes.} use stacked area chart to describe the dynamic fluctuations in block generation across three consensus algorithms: PoW, DPoS, and HL-DPoS, as they interact with varying node counts (500, 1000, 2000, 3000,4000 and 5000) over diverse time spans (10, 20, 30, 40, 50 and 60 minutes). The change in the number of nodes has little impact on the speed of block generation. Among them, DPoS has a similar area height with HL-DPoS in the graph, indicating that HL-DPoS has a similar block generate speed as the DPoS consensus algorithm. 

From Figure \ref{Number of Blocks Changes with the Number of Nodes.}, it can be observed that at 10 minutes, the number of blocks generated by the PoW consensus algorithm is single-digit, represented by the height of the red area on the bottom side of the graph. The change in the number of nodes has little impact on PoW block generation speed. The DPoS and HL-DPoS consensus algorithms generate around 50 blocks, which increase to around 90 at 20 minutes, around 140 at 30 minutes, around 180 at 40 minutes, around 240 at 50 minutes, and around 280 at 60 minutes. The results from the graph indicate that the block generation speed of DPoS and HL-DPoS consensus algorithms remains stable, and the number of blocks generated shows a linear upward trend.

The block generation speed for both DPoS and HL-DPoS are far beyond PoW, because the PoW consensus algorithm requires a lot of computing power, as well as mathematical problems that are dynamically adjusted when obtaining packaging rights. This means that the HL-DPoS consensus algorithm has not reduced the packaging speed while enhancing the security of the DPoS consensus algorithm, maintaining a high level of consensus performance.

\subsection{Centralization emulation}

\begin{figure*}[htbp]
	\centering
	
%

	\subfloat[500 Nodes]{\includegraphics[width=.45\columnwidth]{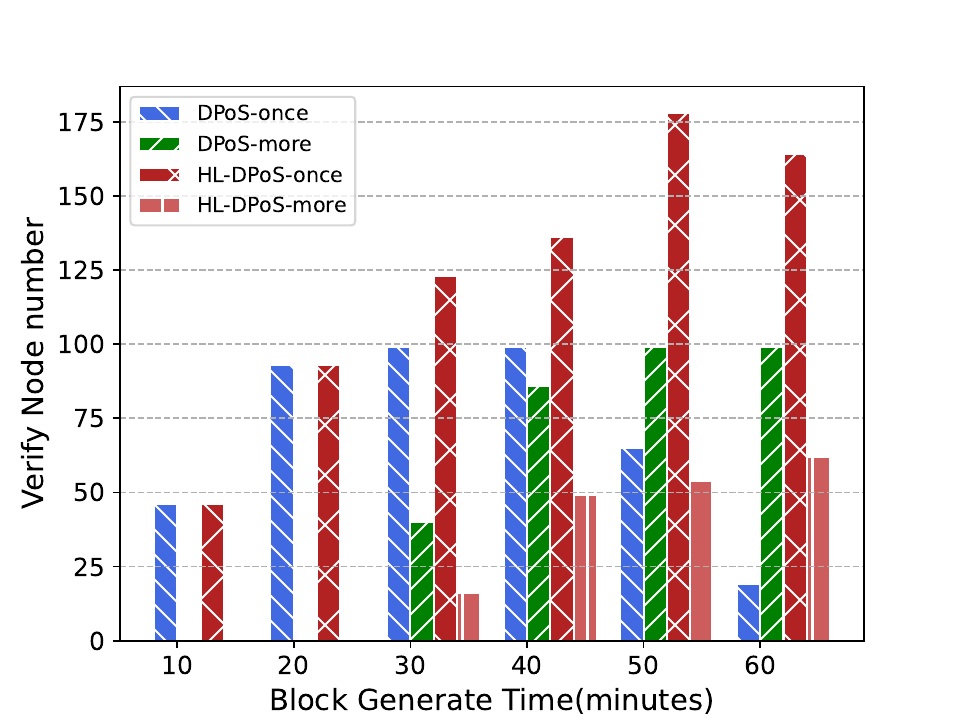}}\hspace{5pt}
        \subfloat[1000 Nodes]{\includegraphics[width=.45\columnwidth]{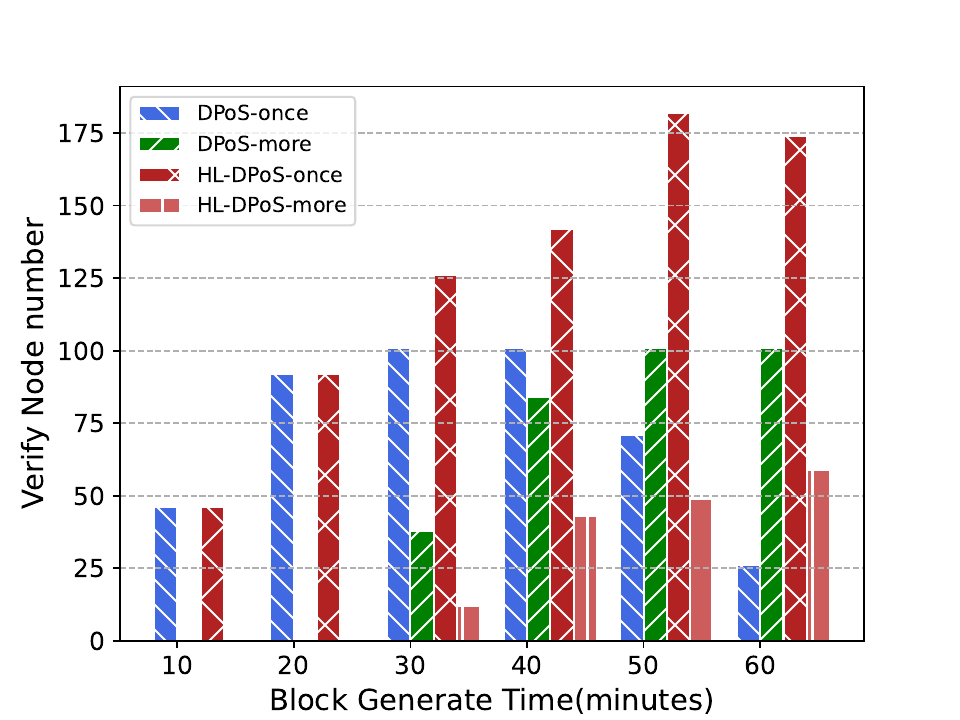}}\\
        
        \subfloat[2000 Nodes]{\includegraphics[width=.45\columnwidth]{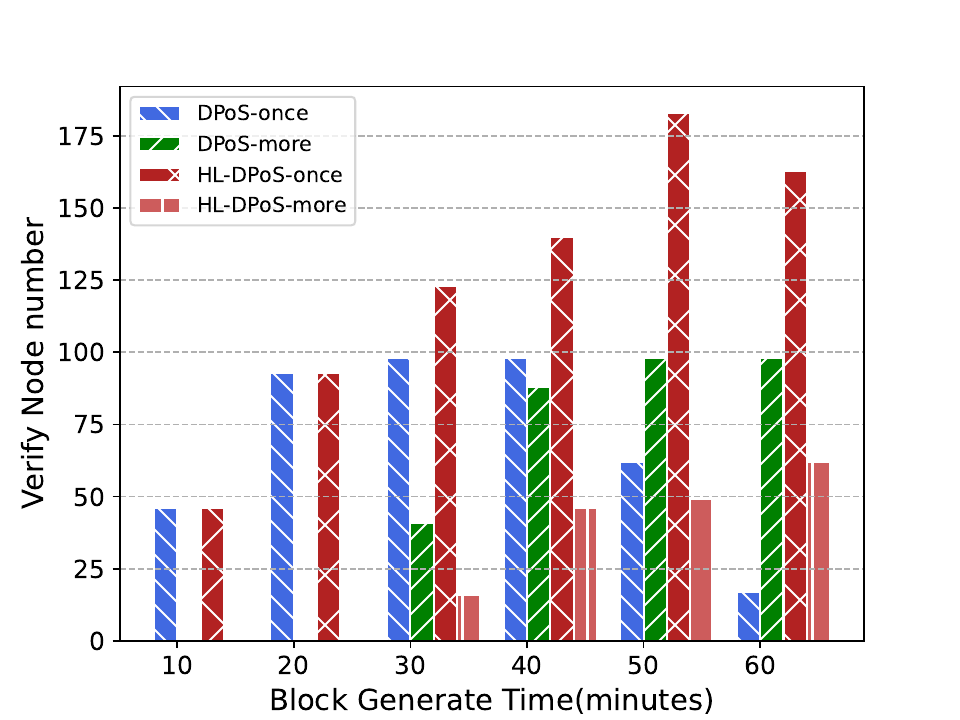}}\hspace{5pt}
	\subfloat[3000 Nodes]{\includegraphics[width=.45\columnwidth]{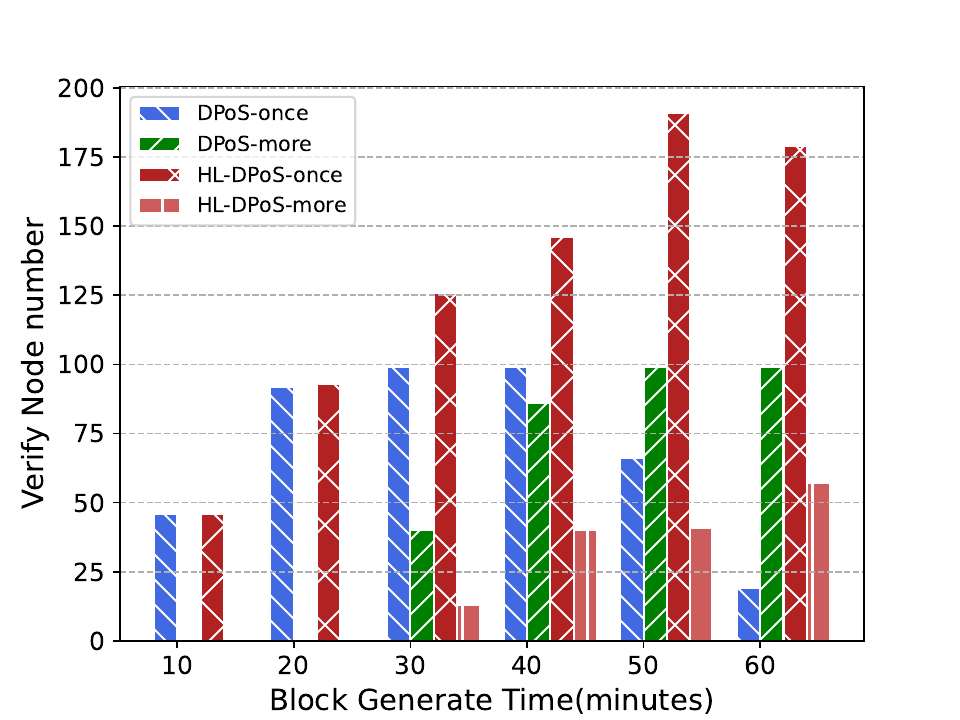}}\\
	
        \subfloat[4000 Nodes]{\includegraphics[width=.45\columnwidth]{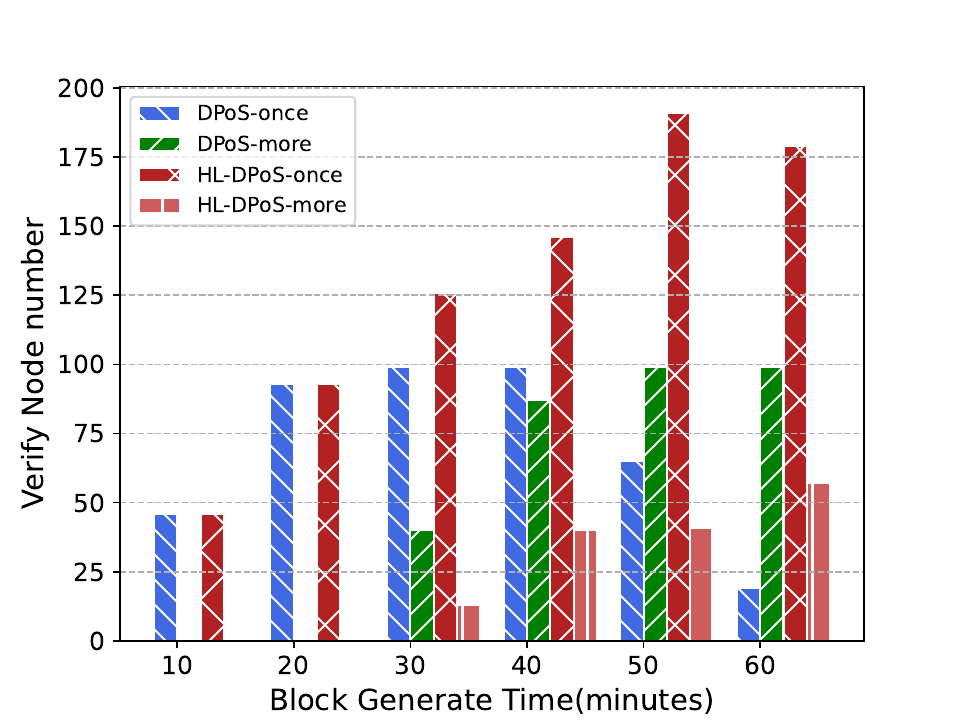}}\hspace{5pt}
        \subfloat[5000 Nodes]{\includegraphics[width=.45\columnwidth]{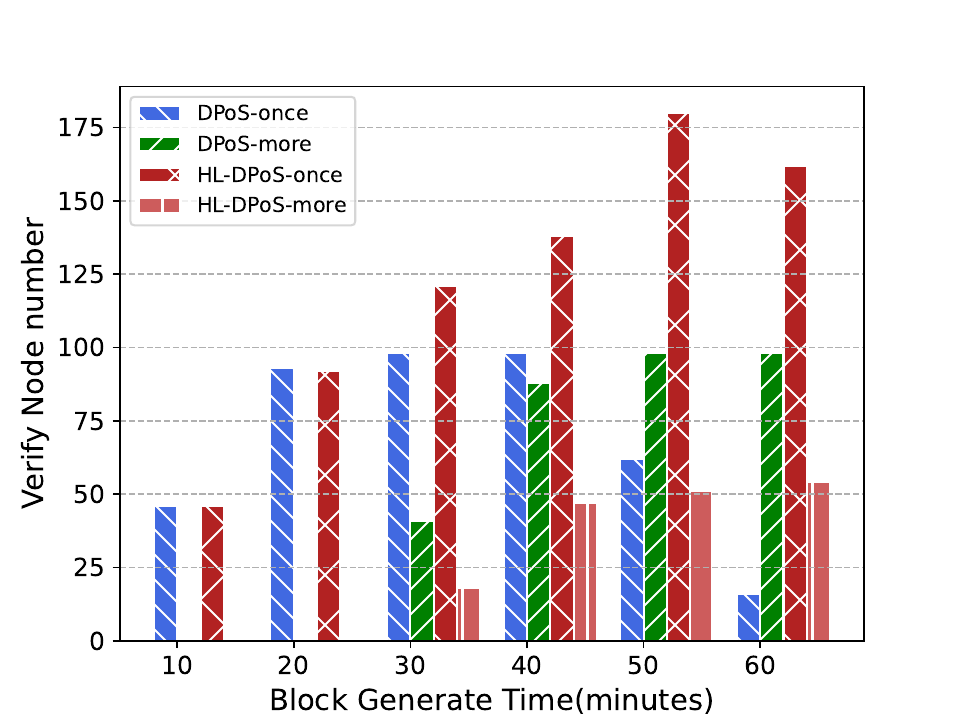}}\\
	\caption{ Times of generating blocks through DPoS and HL-DPoS algorithms, $\psi = 0$, (a) 500 nodes. (b) 1000 nodes. (c) 2000 nodes. (d) 3000 nodes. (e) 4000 nodes. (f) 5000 nodes. }
\label{Times of generating blocks}   
\end{figure*}

Figure \ref{Times of generating blocks} shows the number of witness nodes generated in DPoS and HL-DPoS consensus algorithm with different node sizes (500, 1000, 2000, 3000, 4000, and 5000) at different time intervals (10, 20, 30, 40, 50 and 60 minutes) will be displayed through a bar chart. Then, the difference in the number of times witness nodes selected by the two consensus algorithms enter the packing queue will be compared within the same time range. The specific experimental results are reflected in Figure \ref{Times of generating blocks}. we calculate the witness nodes number of the DPoS and HL-DPoS algorithms in a different time, we denote \textbf{DPoS-once} as the witness nodes in DPoS consensus algorithm packaged one block in a specified time; denotes \textbf{DPoS-more} as the witness nodes in DPoS consensus algorithm packaged more than one block in a specified time; denotes \textbf{HL-DPoS-once} as the witness nodes in HL-DPoS consensus algorithm packaged one block in a specified time; denotes \textbf{HL-DPoS-more} as the witness nodes in HL-DPoS consensus algorithm packaged more than one block in specified time. Among the six subgraphs in Figure \ref{Times of generating blocks}, for the first 20 minutes, no node entered the witness nodes queue more than once, regardless of the different number of nodes in the blockchain. Therefore, both \textbf{DPoS-more} and \textbf{HL-DPoS-more} are 0 during the first 20 minutes and are not displayed in Figure \ref{Times of generating blocks}. \textbf{DPoS-once}, which represents only enters the witness nodes queue once in DPoS consensus algorithm, steadily increases during the first 30 minutes, remains stable between 30 and 40 minutes, and then shows a downward trend after exceeding 40 minutes. However, \textbf{DPoS-more}, which represents the DPoS consensus algorithm and enters the witness nodes queue multiple times, gradually increases from 30 minutes, indicating that more and more nodes repeatedly enter the witness nodes queue. By the 60 minutes, the number of nodes that repeatedly enter the witness nodes queue far exceeds the number of nodes that enter it only once in DPoS. \textbf{HL-DPoS-once}, which represents the HL-DPoS consensus algorithm and only enters the witness nodes queue once, rapidly increases during the first 50 minutes and shows a downward trend at 60 minutes. Meanwhile, \textbf{HL-DPoS-more}, which represents the HL-DPoS consensus algorithm and enters the witness nodes queue multiple times, gradually increases from 30 minutes but is far less than \textbf{HL-DPoS-once} within the 60 minutes time range. From the experimental results, the number of nodes that enter the witness nodes queue only once in the HL-DPoS consensus algorithm is much higher than the number of nodes that enter multiple times, indicating that more nodes can participate in the packing process in the HL-DPoS consensus algorithm. This is likely thanks to the HL-DPoS consensus algorithm uses VRF to group nodes in the blockchain network and selects witness nodes from each group through a combination of election and a random selection, thereby reducing the number of nodes that repeatedly enter the packing queue and lowering the probability of centralization issues arising in the DPoS consensus algorithm.

\begin{figure}[htbp]
    \centering
    \includegraphics[width=.8\textwidth]{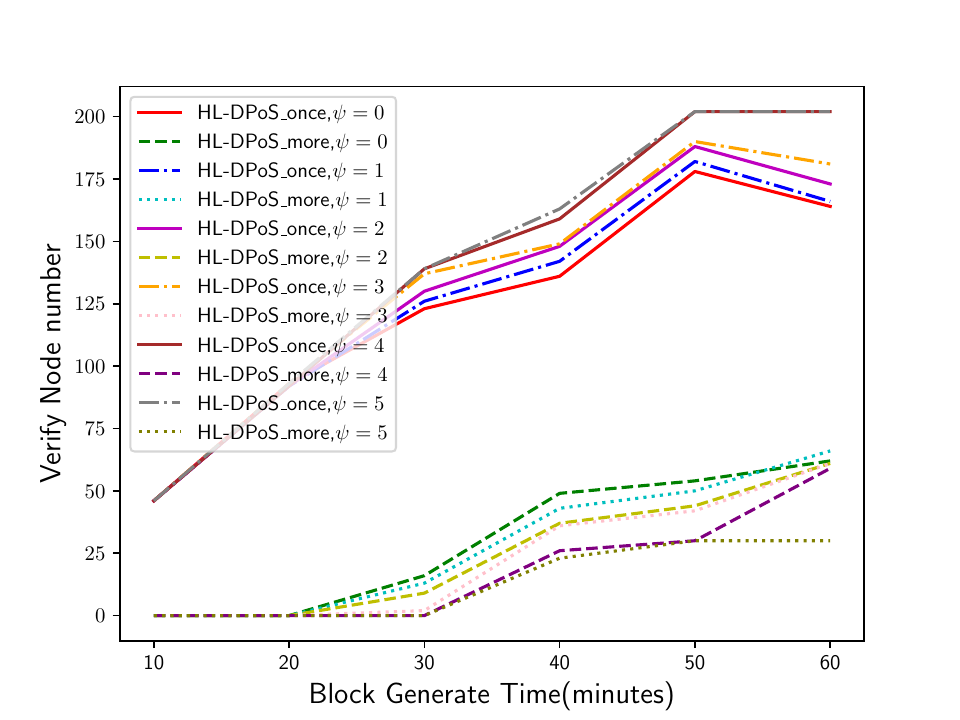}
    \caption{Times of generating blocks through different $psi$ HL-DPoS algorithms, $\psi = 0,1,2,3,4,5$.}
    \label{Times}
    \end{figure} 
 
In Figure \ref{Times}, we emulate different $\psi$ affected nodes repeat enter $\mathbb{PL}$ times. In Figure \ref{Times}, \textbf{HL-DPoS\_once} indicates nodes number that only enter $\mathbb{PL}$ once, \textbf{HL-DPoS\_more} shows nodes number enter $\mathbb{PL}$ more than once, we can find the nodes enter $\mathbb{PL}$ times will decrease while  $\psi$ increases. During the 10 to 50 minute period, the number of nodes entering $\mathbb{PL}$ once continues to increase, far exceeding the number of nodes entering the $\mathbb{PL}$  more than once. Due to the limitation of the total number of nodes, in the final time period, the number of nodes entering the packaging queue once showed a downward trend, while the number of nodes entering the packaging queue more than once showed an upward trend.

\section{CONCLUSION AND FUTURE WORK}

DPoS consensus algorithm has several challenges, including centralization issues and long-range attack issues. 
 We split the nodes into several groups and chose witness nodes by vote and random selection to solve the centralization issue 
 . Proposed longest chain verification mechanism to check if there has a missing transaction and prevent witness nodes' long-range attacks. After algorithm analysis and experimental verification, our proposed HL-DPoS consensus algorithm ensures consensus efficiency and can stop the longest chain merge process in time when there is a lack of transactions that reached consensus, enhancing the algorithm security.

For the consensus algorithm, with the emergence of large language models such as ChatGPT\cite{Kirmani2022ArtificialIS}, the difficulty of attacking blockchain networks is becoming lower and lower,  designing a self-updating and anti-attack blockchain consensus algorithm will be the future research direction.

\section*{Acknowledgement}
This work was supported by the National Natural Science Foundation of China under Grant No. 62272024.






\end{document}